\documentclass[%
reprint,
aps,nofootinbib,
notitlepage,
superscriptaddress,
twocolumn,
nobibnotes,
 amsmath,
 amssymb,
 prl,
]{revtex4-2}

\usepackage{graphicx}% Include figure files
\usepackage{dcolumn}% Align table columns on decimal point
\usepackage{bm}% bold math
\usepackage{braket}
\usepackage{float}
\usepackage{lipsum}
\usepackage{graphicx}% Include figure files
\usepackage{dcolumn}% Align table columns on decimal point
\usepackage{bm}% bold math
\usepackage{natbib}
\usepackage{hyperref}
\usepackage[caption=false]{subfig}
\usepackage{bm}

\hypersetup{
	colorlinks = true,
    linkcolor = Maroon,
    urlcolor  = Maroon,
    citecolor = Maroon
}

\usepackage{microtype}
\usepackage{verbatim}
\usepackage[amssymb]{SIunits}
\usepackage{tabularx}
\usepackage[dvipsnames]{xcolor}
\usepackage{tikz}
\usepackage{orcidlink}
\usepackage[normalem]{ulem}
\usepackage{xcolor}
\usepackage{soul}

\usepackage{verbatim}

\usepackage{color}

\begin{document}

\title{Canonical Hubble-Tension-Resolving Early Dark Energy Cosmologies are Inconsistent with the Lyman-$\alpha$ Forest}

\author{Samuel~Goldstein\,\orcidlink{0000-0003-3155-245X}}
\email{sjg2215@columbia.edu}
\affiliation{Department of Physics, Columbia University, New York, NY 10027, USA}

\author{J.~Colin~Hill\,\orcidlink{0000-0002-9539-0835}}
\affiliation{Department of Physics, Columbia University, New York, NY 10027, USA}

\author{Vid~Ir\v{s}i\v{c}\,\orcidlink{0000-0002-5445-461X}}
\affiliation{Kavli Institute for Cosmology, Madingley Road, Cambridge CB3 0HA, UK}
\affiliation{Cavendish Laboratory, Cambridge, 19 J. J. Thomson Ave, Cambridge CB3
0HE, UK}
\author{Blake D.~Sherwin}%
\affiliation{DAMTP, Centre for Mathematical Sciences, Wilberforce Road, Cambridge CB3 0WA, UK}
\affiliation{Kavli Institute for Cosmology, Madingley Road, Cambridge CB3 0HA, UK}

\date{\today}

%TC:ignore
\begin{abstract}

Current cosmological data exhibit discordance between indirect and some direct inferences of the present-day expansion rate, $H_0$.  Early dark energy (EDE), which briefly increases the cosmic expansion rate prior to recombination, is a leading scenario for resolving this ``Hubble tension'' while preserving a good fit to cosmic microwave background (CMB) data.  However, this comes at the cost of changes in parameters that affect structure formation in the late-time universe, including the spectral index of scalar perturbations, $n_s$.  Here, we present the first constraints on axion-like EDE using data from the Lyman-$\alpha$ forest, i.e., absorption lines imprinted in background quasar spectra by neutral hydrogen gas along the line of sight.  We consider two independent measurements of the one-dimensional Ly$\alpha$ forest flux power spectrum, from the Sloan Digital Sky Survey (SDSS eBOSS) and from the MIKE/HIRES and X-Shooter spectrographs.  We combine these with a baseline dataset comprised of \emph{Planck} CMB data and baryon acoustic oscillation (BAO) measurements.  Combining the eBOSS Ly$\alpha$ data with the CMB and BAO dataset reduces the 95\% confidence level (CL) upper bound on the maximum fractional contribution of EDE to the cosmic energy budget, $f_{\rm EDE}$, from 0.07 to 0.03 and constrains $H_0=67.9_{-0.4}^{+0.4}~{\rm km/s/Mpc}$ (68\% CL), with maximum \textit{a posteriori} value $H_0=67.9~{\rm km/s/Mpc}$. Similar results are obtained for the MIKE/HIRES and X-Shooter Ly$\alpha$ data.  Our Ly$\alpha$-based EDE constraints yield $H_0$ values that are in $>4\sigma$ tension with the SH0ES distance-ladder measurement and are driven by the preference of the Ly$\alpha$ forest data for $n_s$ values lower than those required by EDE cosmologies that fit \emph{Planck} CMB data. Taken at face value, the Ly$\alpha$ forest severely constrains canonical EDE models that could resolve the Hubble tension. 
\end{abstract}

\maketitle
%TC:endignore

\paragraph{Introduction---} The recent direct measurement of the current cosmic expansion rate from the SH0ES Collaboration, $H_0^{\rm SH0ES}=73.04\pm1.04~ {\rm km/s/Mpc}$~\cite{Riess:2021jrx}, is in significant tension with the indirect inference from \emph{Planck} measurements of the cosmic microwave background (CMB), $H_0^{\it Planck}=67.36\pm 0.54~ {\rm km/s/Mpc}$~\cite{Aghanim:2018eyx}, as well as inferences from other CMB experiments~\cite{Aiola2020,SPT-3G:2021eoc} and probes of large-scale structure~(e.g.,~\cite{Schoneberg:2019wmt,Ivanov:2019hqk,Philcox:2020vvt,eBOSS:2020yzd,Philcox:2020xbv, Schoneberg:2022ggi}). Although some local distance ladder probes do not exhibit this discordance~\cite{Freedman:2021ahq,Birrer:2020tax, Wojtak:2022bct}, and thus the origin of this ``Hubble tension'' could be systematic, many new physics models have been proposed as solutions (see \cite{DiValentino:2021izs} for a review). One of the most popular candidates is early dark energy (EDE), in which a scalar field increases the cosmic expansion rate just prior to recombination, before rapidly decaying away so as to not further impact the late universe. EDE decreases the sound horizon at last scattering and thereby increases the value of $H_0$ inferred from CMB analyses~\cite{Bernal:2016gxb, Poulin:2018cxd, Agrawal:2019lmo, Lin:2019qug, Knox:2019rjx, Poulin:2023lkg}. 

Although EDE may resolve the Hubble tension, it does so at the expense of introducing or worsening other tensions when confronted with additional cosmological datasets~\cite{Hill:2020osr, Ivanov:2020ril, DAmico:2020ods, Jedamzik:2020zmd} (see \cite{Smith:2020rxx,Murgia:2020ryi,Simon:2022adh,Herold:2021ksg} for a different viewpoint). As discussed in \cite{Hill:2020osr, Vagnozzi:2021gjh}, EDE models produce an enhanced early integrated Sachs-Wolfe (eISW) effect in the CMB, which must be compensated by larger values of the physical cold dark matter density $\Omega_c h^2$ and the scalar spectral index $n_s$ (as compared to their values in $\Lambda$CDM) in order to fit the CMB data. Conversely, recent analyses of the Lyman-$\alpha$ forest --- absorption features in the spectra of distant quasars due to neutral hydrogen along the line of sight --- prefer values of $n_s$ and $\Omega_m h$ that are lower than those of CMB datasets~\cite{Chabanier:2018rga, Palanque-Delabrouille:2019iyz, Esposito:2022plo}. In this \emph{Letter} we demonstrate that, taken at face value, recent Ly$\alpha$ datasets significantly constrain EDE models.

\paragraph{Model---} We consider EDE composed of a scalar field with an axion-like potential~\cite{Poulin:2018cxd,Smith:2019ihp}, $ V(\phi)= m^2f^2(1-\cos(\phi/f))^n$ where $f$ is the axion decay constant, $m$ is a mass scale, and $n$ is a power-law index. %The dynamics of the field are governed by the Klein-Gordon equation: $\ddot\phi+3H\dot\phi+dV/d\phi=0$ where dots denote derivatives with respect to cosmic time. 
Instead of parametrizing the model in terms of the physical parameters ($m$, $f$) and the initial field value $\phi_i$, we use an effective parametrization defined by the maximum fractional contribution of the EDE field to the cosmic energy budget, $f_{\rm EDE}$, the critical redshift $z_c$ at which the EDE field reaches this contribution, and the initial field displacement $\theta_i\equiv \phi_i/f$~\cite{Poulin:2018cxd,Agrawal:2019lmo}. To be consistent with late-time observables, the EDE field must decay sufficiently rapidly after $z_c$, which requires $n\geq 2$ (thus excluding the standard axion with $n=1$).  In this work, we fix $n=3$, which has been shown to fit current data~\cite{Poulin:2018cxd, Smith:2019ihp}. We compute theoretical predictions using \texttt{CLASS\_EDE}~\cite{Hill:2020osr},\footnote{\url{https://github.com/mwt5345/class_ede}} a modification of the Einstein-Boltzmann code \texttt{CLASS}~\cite{Lesgourgues:2011re, Blas:2011rf} that incorporates EDE dynamics at the background and linear perturbation level.

\paragraph{Datasets---} Our baseline dataset consists of \emph{Planck} 2018 measurements of the CMB temperature and polarization power spectra at small (TTTEEE) and large angular scales (lowl+lowE) \cite{Aghanim:2018eyx, Aghanim:2019ame} and the CMB lensing potential power spectrum \cite{Aghanim:2018oex}, and baryon acoustic oscillation (BAO) measurements from BOSS DR12~\cite{Alam:2016hwk}, the SDSS Main Galaxy Sample~\cite{Ross:2014qpa}, and 6dFGS~\cite{Beutler:2011hx}.

Traditional analyses of Ly$\alpha$ forest flux power spectra interpolate between hydrodynamical simulations to make theory predictions. Given the computational complexity of such simulations, we instead model Ly$\alpha$ forest measurements using a compressed likelihood characterized by the amplitude $\Delta_L^2\equiv k_p^3P_{\rm lin}(k_p,z_p)/(2\pi^2)$ and slope $n_L\equiv \left( d\ln P_{\rm lin}(k, z)/d\ln k \right) |_{(k_p,z_p)}$ of the linear power spectrum $P_{\rm lin}$, both evaluated at a pivot redshift $z_p=3$ and wavenumber $k_p=0.009~{\rm s/km}$ \cite{SDSS:2004aee}. This likelihood is marginalized over astrophysical uncertainties due to baryons. As shown in \cite{Pedersen:2022anu}, $\Delta_L^2$ and $n_L$ contain essentially all of the cosmological information in the Ly$\alpha$ forest flux power spectrum over the range of scales probed by the datasets considered here. 

Our fiducial dataset is derived from the analysis of the 1D Ly$\alpha$ flux power spectrum of SDSS DR14 BOSS and eBOSS quasars \cite{Chabanier:2018rga}, which we refer to as eBOSS Ly$\alpha$. We fit a 2D Gaussian to samples from the $\Delta_L^2$ -- $n_L$ contour shown in Fig.~20 of~\cite{Chabanier:2018rga}. In the Supplemental Material we show that the 2D Gaussian accurately models the contour. The log-likelihood, up to a constant, is
\begin{align}
\log\mathcal{L}= -\frac{1}{2(1-\rho^2)}
\bigg\{ \Delta x^2-2\rho\Delta x\Delta y+\Delta y^2\bigg\},
\end{align}
where $\Delta x\equiv (\Delta_{L}^2-\bar{\Delta}_{L}^{2})/\sigma_{\Delta_L^2}$ and $\Delta y\equiv (n_{L}-\bar{n}_{L})/\sigma_{n_L}$.  Here $(\bar{\Delta}_{L}^{2}, \bar{n}_{L})$ and $(\sigma_{\Delta_L^2}, \sigma_{n_L})$ are the mean and errors of the 2D Gaussian, respectively, and $\rho$ is the correlation coefficient between $\sigma_{\Delta_{L}^2}$ and $\sigma_{n_L}$. Our best-fit parameters describing the eBOSS Ly$\alpha$ dataset are shown in Table~\ref{tab:ly_alpha_gauss_params}.

\begin{table}[!t]
    \centering
    \begin{tabular}{ |c|c|c|c|c|c| } 
     \hline
    Ly$\alpha$ Dataset & $\bar\Delta_{L^2}$ & $\bar{n}_{L}$ & $\sigma_{\Delta_L^2}$ & $\sigma_{n_L}$ & $\rho$ \\ 
    \hline
     eBOSS &  0.310 & -2.340 & 0.020 &  0.006 & 0.512  \\ 
    \hline
     XQ-100/MIKE-HIRES & 0.343 &  -2.388 &  0.033 & 0.021 & 0.694  \\
    \hline
    \end{tabular}
   \caption{Parameter values for the 2D Gaussian compressed likelihoods from the Ly$\alpha$ datasets used in this work.}
   \label{tab:ly_alpha_gauss_params}
\end{table}

The eBOSS Ly$\alpha$ constraints assume a $\Lambda$CDM cosmology with a prior of $H_0=67.3\pm 1.0~{\rm km/s/Mpc}$ and three species of massless neutrinos. The $H_0$ prior and assumptions regarding the neutrino mass have negligible impact on our results because constraints in the $\Delta_L^2$ -- $n_L$ plane are insensitive to the precise value of $H_0$ and $\sum m_\nu$~\cite{Seljak:2006bg, Pedersen:2022anu} (see Appendix A of~\cite{Pedersen:2022anu} and the Supplemental Material of this work for more details). In the Supplemental Material we show that, for the range of scales probed by the Ly$\alpha$ datasets used here, the linear power spectrum for EDE cosmologies that are consistent with the baseline dataset can be mimicked at high precision by a $\Lambda$CDM cosmology; thus the $\Lambda$CDM assumption in the Ly$\alpha$ likelihood has little impact on our results.

 \begin{figure}[!t]
\includegraphics[width=\linewidth]{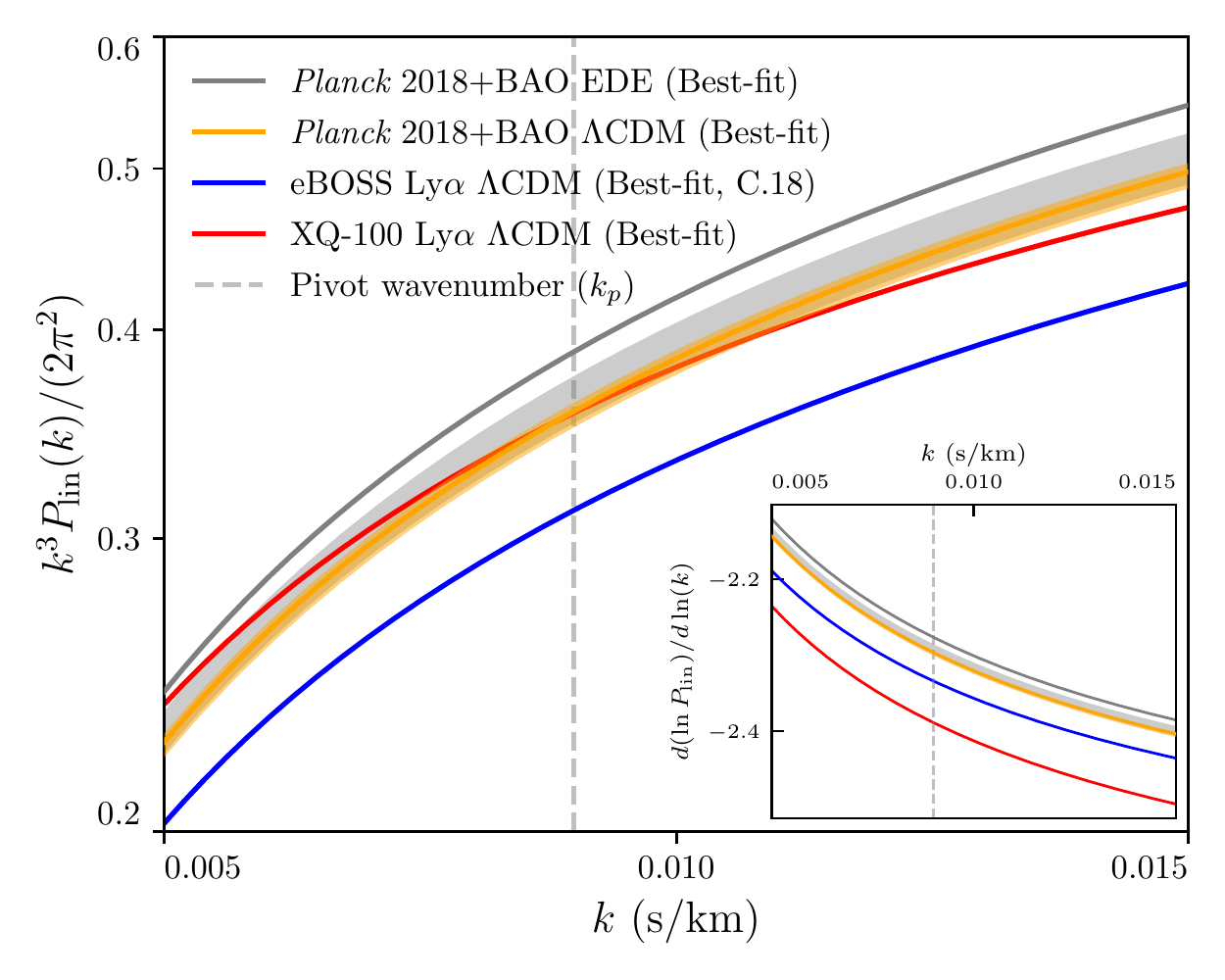}
\caption{Comparison of the best-fit linear matter power spectrum at $z_p=3$ from the EDE (grey) and $\Lambda$CDM (orange) fits to the baseline CMB + BAO dataset with the best-fit $\Lambda$CDM cosmologies for the eBOSS (blue)~\citep{Chabanier:2018rga} and XQ-100 (red) Ly$\alpha$ forest datasets.  
Shaded bands indicate the 68\% CL from our baseline analyses; note that the best-fit EDE model lies outside the 68\% CL due to prior-volume effects. The inset shows the slope $\left( d\ln P_{\rm lin}(k, z)/d\ln k \right) |_{(k_p,z_p)}$ and the vertical line shows the Ly$\alpha$ pivot wavenumber $(k_p=0.009$ s/km).  EDE cosmologies that can resolve the Hubble tension and fit the baseline dataset require an enhanced amplitude and slope near the pivot scale relative to $\Lambda$CDM cosmologies. These requirements, particularly the steeper derivative, are in tension with the Ly$\alpha$ measurements. This figure is for illustrative purposes and thus does not include errors for the Ly$\alpha$ data.}
 \label{fig:ly_alpha_thr_plot}
\end{figure}

\begin{figure*}[!t]
\includegraphics[width=\linewidth]{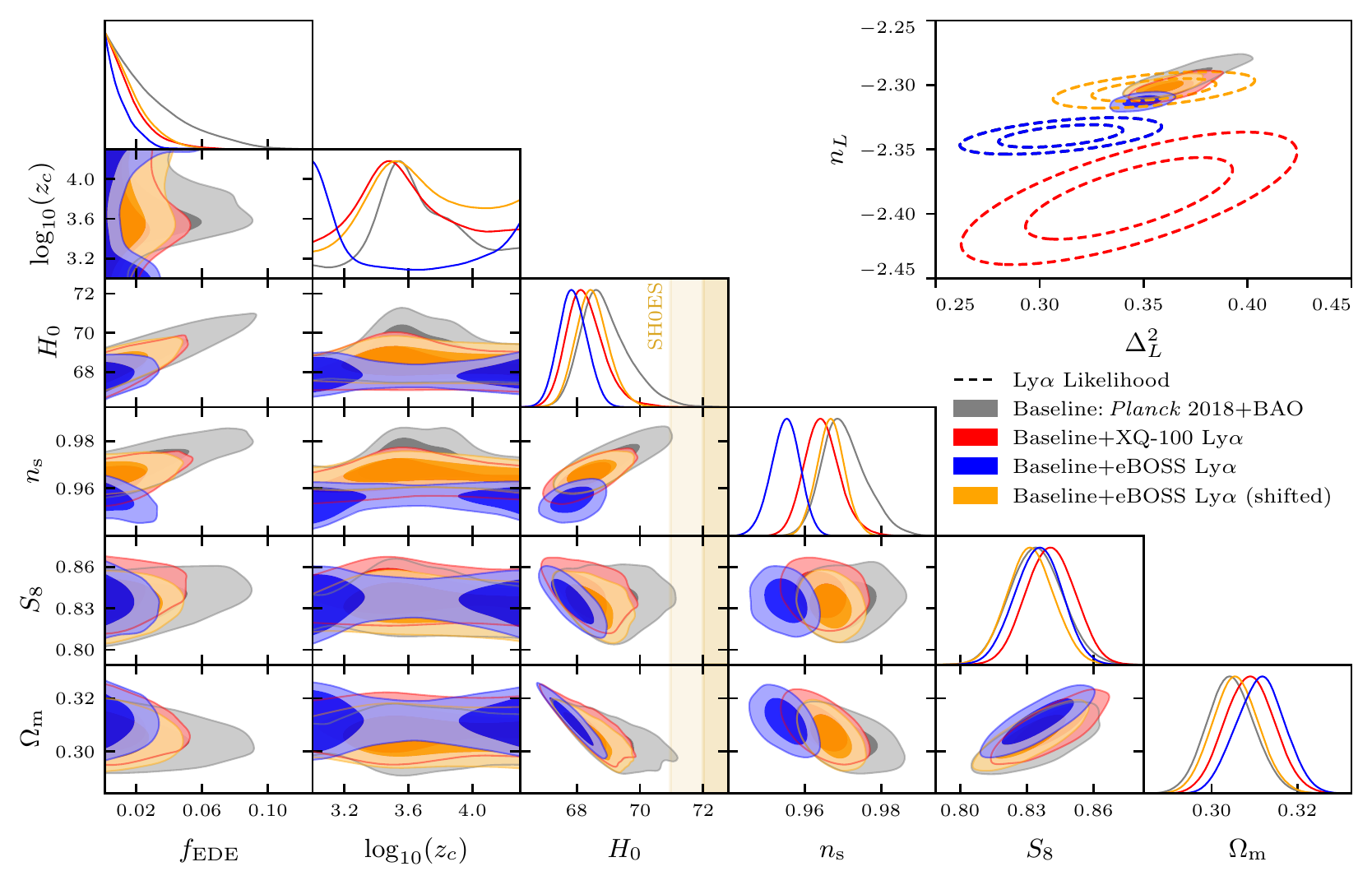}
\caption{Marginalized posteriors for a subset of EDE and $\Lambda$CDM parameters with and without Ly$\alpha$ data. The baseline dataset (grey) consists of \emph{Planck} 2018 high-$\ell$ (TT+TE+EE) and low-$\ell$ (TT+EE) measurements, as well as BOSS DR12, SDSS MGS, and 6dFGS BAO data. Including eBOSS Ly$\alpha$ data (blue) or XQ-100 Ly$\alpha$ data (red) significantly reduces the upper bound on $f_{\rm EDE}$. $H_0$ values that are able to resolve the Hubble tension are strongly excluded by both analyses that include Ly$\alpha$ data. The top right panel shows constraints in the $\Delta_L^2$ -- $n_L$ plane for each of these datasets, and the Ly$\alpha$ likelihoods alone (dashed lines). Although both Ly$\alpha$ likelihoods are in significant tension with the baseline analysis, our conclusions are unchanged even if we artificially shift the center of the eBOSS Ly$\alpha$ likelihood to the posterior mean of the baseline $\Lambda$CDM analysis (orange).}  
 \label{fig:ly_alpha_marg_post_and_likelihood}
\end{figure*}

We also consider measurements from the 1D Ly$\alpha$ forest flux power spectrum of the XQ-100 \cite{Irsic:2017sop} and MIKE/HIRES quasar samples~\cite{Viel:2013fqw}. We fit a 2D Gaussian to the $\Delta_L^2$ -- $n_L$ contour at $z_p=3$ and $k_p=0.009~{\rm s/km}$ derived from the analysis in Appendix A of~\cite{Esposito:2022plo}.\footnote{The XQ100 and MIKE/HIRES samples have a pivot at higher redshifts, but the results can be converted to $z_p=3$ since the rescaling is done in the matter-dominated era.} The details of this likelihood are described in the Supplemental Material.  We refer to this dataset as XQ-100 Ly$\alpha$. Table~\ref{tab:ly_alpha_gauss_params} includes the parameters for this likelihood.

\paragraph{Methodology---} We sample from the EDE parameter posterior distributions using the Markov chain Monte Carlo (MCMC) code \texttt{Cobaya}~\cite{Torrado:2020dgo}.\footnote{\url{https://github.com/CobayaSampler/cobaya}} We compute the effective Ly$\alpha$ parameters $n_L$ and $\Delta_L^2$ from the linear matter power spectrum produced by \texttt{CLASS\_EDE}. We assess convergence using the Gelman-Rubin statistic~\cite{gelman1992} with a tolerance of $|R-1|<0.03.$ We report confidence limits using the credible interval defined in Section IV of \cite{Ivanov:2020ril} and implemented in \texttt{GetDist}~\cite{Lewis:2019xzd}\footnote{\url{https://github.com/cmbant/getdist}} for all parameters, except $f_{\rm EDE}$, for which we quote the 95\% confidence one-tailed upper bound. We determine maximum \emph{a posteriori} (MAP) values using a simulated annealing approach~\cite{Schoneberg:2021qvd} described in the Supplemental Material.

We vary the six $\Lambda$CDM cosmological parameters ($\Omega_bh^2$, $\Omega_ch^2$, $A_s$, $n_s$, $\tau$, $\theta_s$) assuming flat, uninformative priors identical to those in Section 2.1 of~\cite{Aghanim:2018eyx}. We adopt the following uniform priors for the EDE parameters: $f_{\rm EDE}\in [0.001, 0.5],~\log_{10}(z_c)\in[3,4.3],$ and $\theta_i\in[0.1, 3.1].$  The prior on $\log_{10}(z_c)$ is chosen such that we consider only EDE models with dynamics occurring at the epoch to resolve the Hubble tension, which has been shown in previous works (e.g.,~\cite{Poulin:2018cxd,Smith:2019ihp,Hill:2020osr, Hill:2021yec, LaPosta:2021pgm}) to lie in this regime. The prior on $\theta_i$ is chosen to avoid numerical instabilities that arise when initial field values are near $0$ or $\pi.$ For a detailed discussion of priors on EDE parameters, see \cite{Hill:2020osr, Hill:2021yec, LaPosta:2021pgm}. We include all of the recommended nuisance parameters and priors to account for systematic effects in the datasets we consider. In order to be consistent with the eBOSS Ly$\alpha$ likelihood, we assume three species of massless neutrinos.

\begin{table*}[!t]
    \centering
    \begin{tabular}{ |c|c|c|c|c|c|c| } 
      \hline
      & $f_{\rm EDE} $ &  $\log_{10}(z_c)$    &  $n_s$ &   $H_0$ [km/s/Mpc] & $\Omega_ch^2$ \\ 
      \hline
      Baseline     &  $<0.073~(0.072)$ & $3.67^{+0.24}_{-0.30}$ (3.56)& $0.9705^{+0.0046}_{-0.0066}$ (0.9791)& $68.92^{+0.55}_{-0.91}$ (70.33)& $0.1221^{+0.0013}_{-0.0031}$ (0.1267)  \\
      \hline 
      {+SH0ES} &  $0.096^{+0.032}_{-0.026}~(0.114)$  & $3.64^{+0.21}_{-0.16}$ (3.57)& $0.9851^{+0.0065}_{-0.0063}$ (0.9878)& $71.40^{+0.91}_{-0.91}$ (72.01)& $0.1287^{+0.0035}_{-0.0035}$ (0.1311) \\
      \hline
      +eBOSS Ly$\alpha$       &  $<0.028~(0.021)$ &  $3.52^{+0.78}_{-0.52}$ (3.03)& $0.9549^{+0.0039}_{-0.0035}$ (0.9510)& $67.88^{+0.43}_{-0.46}$ (67.81)& $0.1211^{+0.0011}_{-0.0011}$ (0.1213)  \\ 
      \hline
      {+eBOSS Ly$\alpha$+SH0ES} &  $<0.039~(0.026)$ & $3.48^{+0.82}_{-0.48}$ (3.06)& $0.9574^{+0.0044}_{-0.0037}$ (0.9532)& $68.69^{+0.42}_{-0.41}$ (68.66)& $0.1202^{+0.0010}_{-0.0014}$ (0.1203)  \\ 
      \hline
      {+XQ-100 Ly$\alpha$}       &  $<0.041~(0.022)$ & $3.61^{+0.28}_{-0.42}$ (3.53)& $0.9646^{+0.0041}_{-0.0050}$ (0.9666)& $68.28^{+0.47}_{-0.66}$ (68.46)& $0.1216^{+0.0011}_{-0.0018}$ (0.1223)  \\ 
      \hline
      {+XQ-100 Ly$\alpha$+SH0ES}  & $0.060^{+0.025}_{-0.028}~(0.092)$ & $3.55^{+0.06}_{-0.15}$ (3.54)& $0.9750^{+0.0054}_{-0.0060}$ (0.9788)& $70.25^{+0.84}_{-0.84}$ (71.01)& $0.1256^{+0.0032}_{-0.0034}$ (0.1289)\\ 
      \hline
    \end{tabular}

   \caption{Marginalized constraints on cosmological parameters for EDE from the datasets labeled in the first column. For each dataset, we report the posterior mean and the 68\% CL upper and lower limits for all parameters that are detected at $>2\sigma$, otherwise we report the 95\% CL upper limits. MAP values are shown in parentheses. The marginalized constraints for the artificially ``shifted" eBOSS Ly$\alpha$ analysis are included in the Supplemental Material. }
   \label{tab:posterior_param_limits}
\end{table*}

\paragraph{Results---}
To qualitatively demonstrate the incompatibility between the Ly$\alpha$ measurements and EDE cosmologies that are consistent with the baseline dataset and can resolve the Hubble tension, Fig.~\ref{fig:ly_alpha_thr_plot} compares the best-fit linear matter power spectrum and its derivative at the pivot redshift $z_p=3.0$ derived from our baseline $\Lambda$CDM and EDE analyses with that from analyses of the eBOSS~\citep{Chabanier:2018rga} and XQ-100~\citep{Esposito:2022plo} Ly$\alpha$ forest (see the Supplemental Material for the parameter values).  The EDE fit to the baseline dataset prefers enhanced power relative to the $\Lambda$CDM fit at wavenumbers near the pivot scale $k_p = 0.009$ s/km.\footnote{Converting between length and velocity units depends on the cosmology via a factor $H(z_p)/(1+z_p)$.  We find $k_p = 1.00 \, h/{\rm Mpc}$ for the MAP EDE cosmology of the baseline dataset.} 
The best-fit eBOSS Ly$\alpha$ cosmology predicts less power than the baseline $\Lambda$CDM results. Both Ly$\alpha$ datasets prefer a milder slope of the linear matter power spectrum near the pivot wavenumber, as shown in the inset. To compensate for the enhanced eISW effect in the CMB, EDE cosmologies that can preserve the fit to CMB + BAO data and resolve the Hubble tension require an increased amplitude and slope of $P_{\rm lin}(k)$ than that in $\Lambda$CDM cosmologies~\cite{Hill:2020osr,Vagnozzi:2021gjh}. These requirements move precisely against the direction preferred by Ly$\alpha$ data. 

Fig.~\ref{fig:ly_alpha_marg_post_and_likelihood} shows the main results of this \emph{Letter}. The top right panel presents constraints on the effective Ly$\alpha$ parameters $n_L$ and $\Delta_L^2$ for our main analyses, alongside the Ly$\alpha$ likelihoods for the eBOSS and XQ-100 Ly$\alpha$ datasets. The baseline dataset is in significant tension with both Ly$\alpha$ analyses. This tension already exists for $\Lambda$CDM cosmologies, where it is predominantly sourced by the low values of $n_s$ and $\Omega_{c}h^2$ for the eBOSS Ly$\alpha$ analysis~\cite{Chabanier:2018rga} and the low (high) value of $n_s$ ($\sigma_8$) for the XQ-100 analysis~\cite{Esposito:2022plo}. We discuss this tension in the Supplemental Material and emphasize that the direction of the tension is exactly opposite to the parameter shifts necessary for EDE cosmologies that can increase $H_0$. To demonstrate that our conclusions are not a consequence of this tension, we include an analysis with a ``shifted" eBOSS Ly$\alpha$ likelihood centered at the posterior mean of the baseline $\Lambda$CDM analysis $(\bar\Delta_{L^2}, \bar{n}_{L})=(0.355, -2.304).$ 

The remainder of Fig.~\ref{fig:ly_alpha_marg_post_and_likelihood} shows the marginalized posteriors for a subset of the EDE and $\Lambda$CDM parameters. The positive correlation between $f_{\rm EDE}$ and $n_s$ arising from the compensation of the enhanced eISW effect produced in EDE cosmologies is visible in the baseline results. The 95\% CL upper bound on $f_{\rm EDE}$ reduces from 0.07 to 0.03 (0.04) after including eBOSS (XQ-100) Ly$\alpha$ data, strongly excluding models with $f_{\rm EDE} \approx 0.1$, as needed to resolve the Hubble tension~\cite{Poulin:2018cxd}. Including eBOSS data leads to a bimodal $z_c$ posterior. The samples with large values of $\log_{10}(z_c)$ represent scenarios where the EDE field decays too early to have a significant impact on CMB observables.  The samples with low values of $\log_{10}(z_c)$ are associated with changes in the damping physics in EDE cosmologies where $f_{\rm EDE}$ peaks just prior to recombination, as discussed in the Supplemental Material.

Including the shifted eBOSS Ly$\alpha$ likelihood reduces the 95\% CL upper bound on $f_{\rm EDE}$ to 0.04 and provides tighter constraints than XQ-100 on many parameters. These tight constraints are driven by (i) the increased precision of the eBOSS likelihood relative to the XQ-100 likelihood and (ii) the misalignment between the $\Delta_L^2$ -- $n_L$ degeneracy axis of the eBOSS likelihood and that of the baseline analysis. This test illustrates that even if the eBOSS likelihood were not in tension with the baseline dataset, it would still significantly constrain EDE.

Table~\ref{tab:posterior_param_limits} presents marginalized constraints on the cosmological parameters. We also include results with a SH0ES-derived $H_0$ prior, $H_0=73.04\pm1.04~{\rm km/s/Mpc}$~\cite{Riess:2021jrx}. Even with this prior, the $H_0$ posterior for the baseline + eBOSS analysis falls well below the SH0ES measurement. Given the role of prior-volume effects when using MCMC techniques to sample the EDE parameter space~\cite{Smith:2020rxx, Herold:2021ksg, Herold:2022iib}, we compare the posterior mean with the MAP~\cite{Gomez-Valent:2022hkb}. Similar to previous works, we find noticeable disagreement between the posterior mean and MAP for the baseline dataset, indicating that the interpretation of our baseline constraints is sensitive to prior-volume effects. In contrast, for all analyses that include Ly$\alpha$ data, we find excellent agreement between the posterior mean and MAP for all parameters,\footnote{The only exception is $\log_{10}(z_c)$ when including eBOSS Ly$\alpha$, a consequence of the aforementioned bimodality.} suggesting that our Ly$\alpha$ constraints are significantly less sensitive to prior-volume effects than the baseline results. We verify this explicitly in the Supplemental Material using a profile likelihood.

\paragraph{Conclusions---}  In this \emph{Letter} we used two independent measurements of the Ly$\alpha$ forest flux power spectrum to place the first Ly$\alpha$-based constraints on axion-like EDE models. Combining the eBOSS (XQ-100) Ly$\alpha$ data with a baseline dataset comprised of CMB and BAO measurements reduces the 95\% CL upper bound on the maximum fractional contribution of EDE to the cosmic energy budget $f_{\rm EDE}$, from 0.07 to 0.03 (0.04) and constrains $H_0=67.9_{-0.5}^{+0.4}~(68.3_{-0.7}^{+0.5})~{\rm km/s/Mpc}$ at 68\% CL. Our tight constraints are driven by the tension between the low values of $n_s$ preferred by Ly$\alpha$ forest data and the high values of $n_s$ necessary for EDE cosmologies that fit the \emph{Planck} CMB data.

%Taken at face value, the Ly$\alpha$ forest rules out axion-like EDE resolutions to the Hubble tension.
Several caveats arise when applying the compressed likelihoods considered here to constrain EDE cosmologies. First, deriving Ly$\alpha$ constraints on the $\Delta_L^2$ -- $n_L$ plane within the context of EDE cosmologies would require running many hydrodynamical simulations with EDE-based initial conditions (as in, \textit{e.g.},~\cite{Villasenor:2022aiy} for warm dark matter), which is beyond the scope of this work. Second, our conclusions are sensitive to systematics in the measurements of the eBOSS and XQ-100 1D Ly$\alpha$ forest flux power spectra, and the simulations and emulators used to model them. Seeing as the Ly$\alpha$ likelihoods applied in this study are already in tension with the $\Lambda$CDM constraints from CMB and BAO data, and with each other, it is crucial to determine if this tension is a result of systematics~\cite{Fernandez:2023grg}. If the tension is physical, the Ly$\alpha$ forest excludes the canonical EDE model considered in this work as a resolution to the Hubble tension.

There are several ways to extend our analysis. On the theoretical front, it would be useful to develop models that can resolve the Hubble tension without significantly increasing $n_s$, \textit{e.g.}, by invoking EDE couplings to additional fields~\cite{McDonough:2021pdg,Karwal:2021vpk,Lin:2022phm, Berghaus:2022cwf}. In terms of measurements, we made the conservative choice of using only \emph{Planck} CMB and BAO distance measurements, but one could include additional datasets, such as a full-shape likelihood for BOSS galaxy clustering~\cite{Alam:2016hwk, Ivanov:2020ril, DAmico:2020ods, Simon:2022adh}.
Including probes with low values of $S_8$, such as the Dark Energy Survey~\cite{DES_Y3}  would further reduce $f_{\rm EDE}$ and $H_0$ since EDE models exacerbate the ``$S_8$ tension''~\cite{Poulin:2018cxd,Hill:2020osr,Ivanov:2020ril,DAmico:2019fhj}. In light of the preference for non-zero values of $f_{\rm EDE}$ in the recent Atacama Cosmology Telescope results \cite{Hill:2021yec,Poulin:2021bjr,LaPosta:2021pgm}, it would be interesting to repeat our analysis including ACT DR4 data \cite{Aiola2020,ACT:2020frw}.  In the near future, this analysis could be repeated using Ly$\alpha$ forest measurements from the Dark Energy Spectroscopic Instrument~\cite{DESI:2016fyo,DESI:2023pir,DESI:2023xwh,Karacayli:2023afs} and WEAVE~\cite{Jin:2022ddg}, which will provide the largest sample of quasar spectra to date. 

%TC:ignore
\paragraph{Acknowledgements---} We thank Christophe Yeche for providing us with the SDSS DR14 BOSS Ly$\alpha$ compressed likelihood. We are grateful to Simeon Bird, Andreu Font-Ribera, Tanvi Karwal, Pat McDonald, Chris Pedersen, and Oliver Philcox for helpful conversations.  We thank the anonymous referees for helpful suggestions that have greatly improved the manuscript. JCH and BDS thank the organizers of the conference ``Intriguing inconsistencies in the growth of structure over cosmic time'' at the Sexten Center for Astrophysics, where this work originated.  JCH acknowledges support from NSF grant AST-2108536, NASA grant 21-ATP21-0129, DOE grant DE-SC00233966, the Sloan Foundation, and the Simons Foundation. BDS acknowledges support from the European Research Council (ERC) under the European Union’s Horizon 2020 research and innovation programme (Grant agreement No. 851274) and from the Science and Technology Facilities Council (STFC). VI acknowledges support from the Kavli Foundation. We acknowledge computing resources from Columbia University's Shared Research Computing Facility project, which is supported by NIH Research Facility Improvement Grant 1G20RR030893-01, and associated funds from the New York State Empire State Development, Division of Science Technology and Innovation (NYSTAR) Contract C090171, both awarded April 15, 2010. We acknowledge the Texas Advanced Computing Center (TACC) at The University of Texas at Austin for providing HPC resources that have contributed to the research results reported within this paper. Preliminary computations were performed on the HPC system Freya at the Max Planck Computing and Data Facility.  We acknowledge the use of computational resources at the Flatiron Institute.  The Flatiron Institute is supported by the Simons Foundation.

\clearpage

\pagebreak
\bibliographystyle{apsrev4-1}
\bibliography{biblio}

\providecommand{\noopsort}[1]{}\providecommand{\singleletter}[1]{#1}%
\begin{thebibliography}{81}%
\makeatletter
\providecommand \@ifxundefined [1]{%
 \@ifx{#1\undefined}
}%
\providecommand \@ifnum [1]{%
 \ifnum #1\expandafter \@firstoftwo
 \else \expandafter \@secondoftwo
 \fi
}%
\providecommand \@ifx [1]{%
 \ifx #1\expandafter \@firstoftwo
 \else \expandafter \@secondoftwo
 \fi
}%
\providecommand \natexlab [1]{#1}%
\providecommand \enquote  [1]{``#1''}%
\providecommand \bibnamefont  [1]{#1}%
\providecommand \bibfnamefont [1]{#1}%
\providecommand \citenamefont [1]{#1}%
\providecommand \href@noop [0]{\@secondoftwo}%
\providecommand \href [0]{\begingroup \@sanitize@url \@href}%
\providecommand \@href[1]{\@@startlink{#1}\@@href}%
\providecommand \@@href[1]{\endgroup#1\@@endlink}%
\providecommand \@sanitize@url [0]{\catcode `\\12\catcode `\$12\catcode
  `\&12\catcode `\#12\catcode `\^12\catcode `\_12\catcode `\%12\relax}%
\providecommand \@@startlink[1]{}%
\providecommand \@@endlink[0]{}%
\providecommand \url  [0]{\begingroup\@sanitize@url \@url }%
\providecommand \@url [1]{\endgroup\@href {#1}{\urlprefix }}%
\providecommand \urlprefix  [0]{URL }%
\providecommand \Eprint [0]{\href }%
\providecommand \doibase [0]{http://dx.doi.org/}%
\providecommand \selectlanguage [0]{\@gobble}%
\providecommand \bibinfo  [0]{\@secondoftwo}%
\providecommand \bibfield  [0]{\@secondoftwo}%
\providecommand \translation [1]{[#1]}%
\providecommand \BibitemOpen [0]{}%
\providecommand \bibitemStop [0]{}%
\providecommand \bibitemNoStop [0]{.\EOS\space}%
\providecommand \EOS [0]{\spacefactor3000\relax}%
\providecommand \BibitemShut  [1]{\csname bibitem#1\endcsname}%
\let\auto@bib@innerbib\@empty
%</preamble>
\bibitem [{\citenamefont {Riess}\ \emph {et~al.}(2022)\citenamefont {Riess}
  \emph {et~al.}}]{Riess:2021jrx}%
  \BibitemOpen
  \bibfield  {author} {\bibinfo {author} {\bibfnamefont {A.~G.}\ \bibnamefont
  {Riess}} \emph {et~al.},\ }\href {\doibase 10.3847/2041-8213/ac5c5b}
  {\bibfield  {journal} {\bibinfo  {journal} {Astrophys. J. Lett.}\ }\textbf
  {\bibinfo {volume} {934}},\ \bibinfo {pages} {L7} (\bibinfo {year} {2022})},\
  \Eprint {http://arxiv.org/abs/2112.04510} {arXiv:2112.04510 [astro-ph.CO]}
  \BibitemShut {NoStop}%
\bibitem [{\citenamefont {Aghanim}\ \emph
  {et~al.}(2020{\natexlab{a}})\citenamefont {Aghanim} \emph
  {et~al.}}]{Aghanim:2018eyx}%
  \BibitemOpen
  \bibfield  {author} {\bibinfo {author} {\bibfnamefont {N.}~\bibnamefont
  {Aghanim}} \emph {et~al.} (\bibinfo {collaboration} {Planck}),\ }\href
  {\doibase 10.1051/0004-6361/201833910} {\bibfield  {journal} {\bibinfo
  {journal} {Astron. Astrophys.}\ }\textbf {\bibinfo {volume} {641}},\ \bibinfo
  {pages} {A6} (\bibinfo {year} {2020}{\natexlab{a}})},\ \Eprint
  {http://arxiv.org/abs/1807.06209} {arXiv:1807.06209 [astro-ph.CO]}
  \BibitemShut {NoStop}%
\bibitem [{\citenamefont {Aiola}\ \emph {et~al.}(2020)\citenamefont {Aiola}
  \emph {et~al.}}]{Aiola2020}%
  \BibitemOpen
  \bibfield  {author} {\bibinfo {author} {\bibfnamefont {S.}~\bibnamefont
  {Aiola}} \emph {et~al.} (\bibinfo {collaboration} {ACT}),\ }\href {\doibase
  10.1088/1475-7516/2020/12/047} {\bibfield  {journal} {\bibinfo  {journal}
  {JCAP}\ }\textbf {\bibinfo {volume} {12}},\ \bibinfo {pages} {047} (\bibinfo
  {year} {2020})},\ \Eprint {http://arxiv.org/abs/2007.07288} {arXiv:2007.07288
  [astro-ph.CO]} \BibitemShut {NoStop}%
\bibitem [{\citenamefont {Dutcher}\ \emph {et~al.}(2021)\citenamefont {Dutcher}
  \emph {et~al.}}]{SPT-3G:2021eoc}%
  \BibitemOpen
  \bibfield  {author} {\bibinfo {author} {\bibfnamefont {D.}~\bibnamefont
  {Dutcher}} \emph {et~al.} (\bibinfo {collaboration} {SPT-3G}),\ }\href
  {\doibase 10.1103/PhysRevD.104.022003} {\bibfield  {journal} {\bibinfo
  {journal} {Phys. Rev. D}\ }\textbf {\bibinfo {volume} {104}},\ \bibinfo
  {pages} {022003} (\bibinfo {year} {2021})},\ \Eprint
  {http://arxiv.org/abs/2101.01684} {arXiv:2101.01684 [astro-ph.CO]}
  \BibitemShut {NoStop}%
\bibitem [{\citenamefont {Sch\"oneberg}\ \emph {et~al.}(2019)\citenamefont
  {Sch\"oneberg}, \citenamefont {Lesgourgues},\ and\ \citenamefont
  {Hooper}}]{Schoneberg:2019wmt}%
  \BibitemOpen
  \bibfield  {author} {\bibinfo {author} {\bibfnamefont {N.}~\bibnamefont
  {Sch\"oneberg}}, \bibinfo {author} {\bibfnamefont {J.}~\bibnamefont
  {Lesgourgues}}, \ and\ \bibinfo {author} {\bibfnamefont {D.~C.}\ \bibnamefont
  {Hooper}},\ }\href {\doibase 10.1088/1475-7516/2019/10/029} {\bibfield
  {journal} {\bibinfo  {journal} {JCAP}\ }\textbf {\bibinfo {volume} {10}},\
  \bibinfo {pages} {029} (\bibinfo {year} {2019})},\ \Eprint
  {http://arxiv.org/abs/1907.11594} {arXiv:1907.11594 [astro-ph.CO]}
  \BibitemShut {NoStop}%
\bibitem [{\citenamefont {Ivanov}\ \emph
  {et~al.}(2020{\natexlab{a}})\citenamefont {Ivanov}, \citenamefont
  {Simonovi\'c},\ and\ \citenamefont {Zaldarriaga}}]{Ivanov:2019hqk}%
  \BibitemOpen
  \bibfield  {author} {\bibinfo {author} {\bibfnamefont {M.~M.}\ \bibnamefont
  {Ivanov}}, \bibinfo {author} {\bibfnamefont {M.}~\bibnamefont {Simonovi\'c}},
  \ and\ \bibinfo {author} {\bibfnamefont {M.}~\bibnamefont {Zaldarriaga}},\
  }\href {\doibase 10.1103/PhysRevD.101.083504} {\bibfield  {journal} {\bibinfo
   {journal} {Phys. Rev. D}\ }\textbf {\bibinfo {volume} {101}},\ \bibinfo
  {pages} {083504} (\bibinfo {year} {2020}{\natexlab{a}})},\ \Eprint
  {http://arxiv.org/abs/1912.08208} {arXiv:1912.08208 [astro-ph.CO]}
  \BibitemShut {NoStop}%
\bibitem [{\citenamefont {Philcox}\ \emph {et~al.}(2020)\citenamefont
  {Philcox}, \citenamefont {Ivanov}, \citenamefont {Simonovi\'c},\ and\
  \citenamefont {Zaldarriaga}}]{Philcox:2020vvt}%
  \BibitemOpen
  \bibfield  {author} {\bibinfo {author} {\bibfnamefont {O.~H.~E.}\
  \bibnamefont {Philcox}}, \bibinfo {author} {\bibfnamefont {M.~M.}\
  \bibnamefont {Ivanov}}, \bibinfo {author} {\bibfnamefont {M.}~\bibnamefont
  {Simonovi\'c}}, \ and\ \bibinfo {author} {\bibfnamefont {M.}~\bibnamefont
  {Zaldarriaga}},\ }\href {\doibase 10.1088/1475-7516/2020/05/032} {\bibfield
  {journal} {\bibinfo  {journal} {JCAP}\ }\textbf {\bibinfo {volume} {05}},\
  \bibinfo {pages} {032} (\bibinfo {year} {2020})},\ \Eprint
  {http://arxiv.org/abs/2002.04035} {arXiv:2002.04035 [astro-ph.CO]}
  \BibitemShut {NoStop}%
\bibitem [{\citenamefont {Alam}\ \emph {et~al.}(2021)\citenamefont {Alam} \emph
  {et~al.}}]{eBOSS:2020yzd}%
  \BibitemOpen
  \bibfield  {author} {\bibinfo {author} {\bibfnamefont {S.}~\bibnamefont
  {Alam}} \emph {et~al.} (\bibinfo {collaboration} {eBOSS}),\ }\href {\doibase
  10.1103/PhysRevD.103.083533} {\bibfield  {journal} {\bibinfo  {journal}
  {Phys. Rev. D}\ }\textbf {\bibinfo {volume} {103}},\ \bibinfo {pages}
  {083533} (\bibinfo {year} {2021})},\ \Eprint
  {http://arxiv.org/abs/2007.08991} {arXiv:2007.08991 [astro-ph.CO]}
  \BibitemShut {NoStop}%
\bibitem [{\citenamefont {Philcox}\ \emph {et~al.}(2021)\citenamefont
  {Philcox}, \citenamefont {Sherwin}, \citenamefont {Farren},\ and\
  \citenamefont {Baxter}}]{Philcox:2020xbv}%
  \BibitemOpen
  \bibfield  {author} {\bibinfo {author} {\bibfnamefont {O.~H.~E.}\
  \bibnamefont {Philcox}}, \bibinfo {author} {\bibfnamefont {B.~D.}\
  \bibnamefont {Sherwin}}, \bibinfo {author} {\bibfnamefont {G.~S.}\
  \bibnamefont {Farren}}, \ and\ \bibinfo {author} {\bibfnamefont {E.~J.}\
  \bibnamefont {Baxter}},\ }\href {\doibase 10.1103/PhysRevD.103.023538}
  {\bibfield  {journal} {\bibinfo  {journal} {Phys. Rev. D}\ }\textbf {\bibinfo
  {volume} {103}},\ \bibinfo {pages} {023538} (\bibinfo {year} {2021})},\
  \Eprint {http://arxiv.org/abs/2008.08084} {arXiv:2008.08084 [astro-ph.CO]}
  \BibitemShut {NoStop}%
\bibitem [{\citenamefont {Sch\"oneberg}\ \emph
  {et~al.}(2022{\natexlab{a}})\citenamefont {Sch\"oneberg}, \citenamefont
  {Verde}, \citenamefont {Gil-Mar\'\i{}n},\ and\ \citenamefont
  {Brieden}}]{Schoneberg:2022ggi}%
  \BibitemOpen
  \bibfield  {author} {\bibinfo {author} {\bibfnamefont {N.}~\bibnamefont
  {Sch\"oneberg}}, \bibinfo {author} {\bibfnamefont {L.}~\bibnamefont {Verde}},
  \bibinfo {author} {\bibfnamefont {H.}~\bibnamefont {Gil-Mar\'\i{}n}}, \ and\
  \bibinfo {author} {\bibfnamefont {S.}~\bibnamefont {Brieden}},\ }\href
  {\doibase 10.1088/1475-7516/2022/11/039} {\bibfield  {journal} {\bibinfo
  {journal} {JCAP}\ }\textbf {\bibinfo {volume} {11}},\ \bibinfo {pages} {039}
  (\bibinfo {year} {2022}{\natexlab{a}})},\ \Eprint
  {http://arxiv.org/abs/2209.14330} {arXiv:2209.14330 [astro-ph.CO]}
  \BibitemShut {NoStop}%
\bibitem [{\citenamefont {Freedman}(2021)}]{Freedman:2021ahq}%
  \BibitemOpen
  \bibfield  {author} {\bibinfo {author} {\bibfnamefont {W.~L.}\ \bibnamefont
  {Freedman}},\ }\href {\doibase 10.3847/1538-4357/ac0e95} {\bibfield
  {journal} {\bibinfo  {journal} {Astrophys. J.}\ }\textbf {\bibinfo {volume}
  {919}},\ \bibinfo {pages} {16} (\bibinfo {year} {2021})},\ \Eprint
  {http://arxiv.org/abs/2106.15656} {arXiv:2106.15656 [astro-ph.CO]}
  \BibitemShut {NoStop}%
\bibitem [{\citenamefont {Birrer}\ \emph {et~al.}(2020)\citenamefont {Birrer}
  \emph {et~al.}}]{Birrer:2020tax}%
  \BibitemOpen
  \bibfield  {author} {\bibinfo {author} {\bibfnamefont {S.}~\bibnamefont
  {Birrer}} \emph {et~al.},\ }\href {\doibase 10.1051/0004-6361/202038861}
  {\bibfield  {journal} {\bibinfo  {journal} {Astron. Astrophys.}\ }\textbf
  {\bibinfo {volume} {643}},\ \bibinfo {pages} {A165} (\bibinfo {year}
  {2020})},\ \Eprint {http://arxiv.org/abs/2007.02941} {arXiv:2007.02941
  [astro-ph.CO]} \BibitemShut {NoStop}%
\bibitem [{\citenamefont {Wojtak}\ and\ \citenamefont
  {Hjorth}(2022)}]{Wojtak:2022bct}%
  \BibitemOpen
  \bibfield  {author} {\bibinfo {author} {\bibfnamefont {R.}~\bibnamefont
  {Wojtak}}\ and\ \bibinfo {author} {\bibfnamefont {J.}~\bibnamefont
  {Hjorth}},\ }\href {\doibase 10.1093/mnras/stac1878} {\bibfield  {journal}
  {\bibinfo  {journal} {Mon. Not. Roy. Astron. Soc.}\ }\textbf {\bibinfo
  {volume} {515}},\ \bibinfo {pages} {2790} (\bibinfo {year} {2022})},\ \Eprint
  {http://arxiv.org/abs/2206.08160} {arXiv:2206.08160 [astro-ph.CO]}
  \BibitemShut {NoStop}%
\bibitem [{\citenamefont {Di~Valentino}\ \emph {et~al.}(2021)\citenamefont
  {Di~Valentino}, \citenamefont {Mena}, \citenamefont {Pan}, \citenamefont
  {Visinelli}, \citenamefont {Yang}, \citenamefont {Melchiorri}, \citenamefont
  {Mota}, \citenamefont {Riess},\ and\ \citenamefont
  {Silk}}]{DiValentino:2021izs}%
  \BibitemOpen
  \bibfield  {author} {\bibinfo {author} {\bibfnamefont {E.}~\bibnamefont
  {Di~Valentino}}, \bibinfo {author} {\bibfnamefont {O.}~\bibnamefont {Mena}},
  \bibinfo {author} {\bibfnamefont {S.}~\bibnamefont {Pan}}, \bibinfo {author}
  {\bibfnamefont {L.}~\bibnamefont {Visinelli}}, \bibinfo {author}
  {\bibfnamefont {W.}~\bibnamefont {Yang}}, \bibinfo {author} {\bibfnamefont
  {A.}~\bibnamefont {Melchiorri}}, \bibinfo {author} {\bibfnamefont {D.~F.}\
  \bibnamefont {Mota}}, \bibinfo {author} {\bibfnamefont {A.~G.}\ \bibnamefont
  {Riess}}, \ and\ \bibinfo {author} {\bibfnamefont {J.}~\bibnamefont {Silk}},\
  }\href {\doibase 10.1088/1361-6382/ac086d} {\bibfield  {journal} {\bibinfo
  {journal} {Class. Quant. Grav.}\ }\textbf {\bibinfo {volume} {38}},\ \bibinfo
  {pages} {153001} (\bibinfo {year} {2021})},\ \Eprint
  {http://arxiv.org/abs/2103.01183} {arXiv:2103.01183 [astro-ph.CO]}
  \BibitemShut {NoStop}%
\bibitem [{\citenamefont {Bernal}\ \emph {et~al.}(2016)\citenamefont {Bernal},
  \citenamefont {Verde},\ and\ \citenamefont {Riess}}]{Bernal:2016gxb}%
  \BibitemOpen
  \bibfield  {author} {\bibinfo {author} {\bibfnamefont {J.~L.}\ \bibnamefont
  {Bernal}}, \bibinfo {author} {\bibfnamefont {L.}~\bibnamefont {Verde}}, \
  and\ \bibinfo {author} {\bibfnamefont {A.~G.}\ \bibnamefont {Riess}},\ }\href
  {\doibase 10.1088/1475-7516/2016/10/019} {\bibfield  {journal} {\bibinfo
  {journal} {JCAP}\ }\textbf {\bibinfo {volume} {10}},\ \bibinfo {pages} {019}
  (\bibinfo {year} {2016})},\ \Eprint {http://arxiv.org/abs/1607.05617}
  {arXiv:1607.05617 [astro-ph.CO]} \BibitemShut {NoStop}%
\bibitem [{\citenamefont {Poulin}\ \emph {et~al.}(2019)\citenamefont {Poulin},
  \citenamefont {Smith}, \citenamefont {Karwal},\ and\ \citenamefont
  {Kamionkowski}}]{Poulin:2018cxd}%
  \BibitemOpen
  \bibfield  {author} {\bibinfo {author} {\bibfnamefont {V.}~\bibnamefont
  {Poulin}}, \bibinfo {author} {\bibfnamefont {T.~L.}\ \bibnamefont {Smith}},
  \bibinfo {author} {\bibfnamefont {T.}~\bibnamefont {Karwal}}, \ and\ \bibinfo
  {author} {\bibfnamefont {M.}~\bibnamefont {Kamionkowski}},\ }\href {\doibase
  10.1103/PhysRevLett.122.221301} {\bibfield  {journal} {\bibinfo  {journal}
  {Phys. Rev. Lett.}\ }\textbf {\bibinfo {volume} {122}},\ \bibinfo {pages}
  {221301} (\bibinfo {year} {2019})},\ \Eprint
  {http://arxiv.org/abs/1811.04083} {arXiv:1811.04083 [astro-ph.CO]}
  \BibitemShut {NoStop}%
\bibitem [{\citenamefont {Agrawal}\ \emph {et~al.}(2019)\citenamefont
  {Agrawal}, \citenamefont {Cyr-Racine}, \citenamefont {Pinner},\ and\
  \citenamefont {Randall}}]{Agrawal:2019lmo}%
  \BibitemOpen
  \bibfield  {author} {\bibinfo {author} {\bibfnamefont {P.}~\bibnamefont
  {Agrawal}}, \bibinfo {author} {\bibfnamefont {F.-Y.}\ \bibnamefont
  {Cyr-Racine}}, \bibinfo {author} {\bibfnamefont {D.}~\bibnamefont {Pinner}},
  \ and\ \bibinfo {author} {\bibfnamefont {L.}~\bibnamefont {Randall}},\
  }\href@noop {} {\  (\bibinfo {year} {2019})},\ \Eprint
  {http://arxiv.org/abs/1904.01016} {arXiv:1904.01016 [astro-ph.CO]}
  \BibitemShut {NoStop}%
%%CITATION = ARXIV:1904.01016;%%
\bibitem [{\citenamefont {Lin}\ \emph {et~al.}(2019)\citenamefont {Lin},
  \citenamefont {Benevento}, \citenamefont {Hu},\ and\ \citenamefont
  {Raveri}}]{Lin:2019qug}%
  \BibitemOpen
  \bibfield  {author} {\bibinfo {author} {\bibfnamefont {M.-X.}\ \bibnamefont
  {Lin}}, \bibinfo {author} {\bibfnamefont {G.}~\bibnamefont {Benevento}},
  \bibinfo {author} {\bibfnamefont {W.}~\bibnamefont {Hu}}, \ and\ \bibinfo
  {author} {\bibfnamefont {M.}~\bibnamefont {Raveri}},\ }\href {\doibase
  10.1103/PhysRevD.100.063542} {\bibfield  {journal} {\bibinfo  {journal}
  {Phys. Rev.}\ }\textbf {\bibinfo {volume} {D100}},\ \bibinfo {pages} {063542}
  (\bibinfo {year} {2019})},\ \Eprint {http://arxiv.org/abs/1905.12618}
  {arXiv:1905.12618 [astro-ph.CO]} \BibitemShut {NoStop}%
%%CITATION = ARXIV:1905.12618;%%
\bibitem [{\citenamefont {Knox}\ and\ \citenamefont
  {Millea}(2020)}]{Knox:2019rjx}%
  \BibitemOpen
  \bibfield  {author} {\bibinfo {author} {\bibfnamefont {L.}~\bibnamefont
  {Knox}}\ and\ \bibinfo {author} {\bibfnamefont {M.}~\bibnamefont {Millea}},\
  }\href {\doibase 10.1103/PhysRevD.101.043533} {\bibfield  {journal} {\bibinfo
   {journal} {Phys. Rev.}\ }\textbf {\bibinfo {volume} {D101}},\ \bibinfo
  {pages} {043533} (\bibinfo {year} {2020})},\ \Eprint
  {http://arxiv.org/abs/1908.03663} {arXiv:1908.03663 [astro-ph.CO]}
  \BibitemShut {NoStop}%
%%CITATION = ARXIV:1908.03663;%%
\bibitem [{\citenamefont {Poulin}\ \emph {et~al.}(2023)\citenamefont {Poulin},
  \citenamefont {Smith},\ and\ \citenamefont {Karwal}}]{Poulin:2023lkg}%
  \BibitemOpen
  \bibfield  {author} {\bibinfo {author} {\bibfnamefont {V.}~\bibnamefont
  {Poulin}}, \bibinfo {author} {\bibfnamefont {T.~L.}\ \bibnamefont {Smith}}, \
  and\ \bibinfo {author} {\bibfnamefont {T.}~\bibnamefont {Karwal}},\
  }\href@noop {} {\  (\bibinfo {year} {2023})},\ \Eprint
  {http://arxiv.org/abs/2302.09032} {arXiv:2302.09032 [astro-ph.CO]}
  \BibitemShut {NoStop}%
\bibitem [{\citenamefont {Hill}\ \emph {et~al.}(2020)\citenamefont {Hill},
  \citenamefont {McDonough}, \citenamefont {Toomey},\ and\ \citenamefont
  {Alexander}}]{Hill:2020osr}%
  \BibitemOpen
  \bibfield  {author} {\bibinfo {author} {\bibfnamefont {J.~C.}\ \bibnamefont
  {Hill}}, \bibinfo {author} {\bibfnamefont {E.}~\bibnamefont {McDonough}},
  \bibinfo {author} {\bibfnamefont {M.~W.}\ \bibnamefont {Toomey}}, \ and\
  \bibinfo {author} {\bibfnamefont {S.}~\bibnamefont {Alexander}},\ }\href
  {\doibase 10.1103/PhysRevD.102.043507} {\bibfield  {journal} {\bibinfo
  {journal} {Phys. Rev. D}\ }\textbf {\bibinfo {volume} {102}},\ \bibinfo
  {pages} {043507} (\bibinfo {year} {2020})},\ \Eprint
  {http://arxiv.org/abs/2003.07355} {arXiv:2003.07355 [astro-ph.CO]}
  \BibitemShut {NoStop}%
\bibitem [{\citenamefont {Ivanov}\ \emph
  {et~al.}(2020{\natexlab{b}})\citenamefont {Ivanov}, \citenamefont
  {McDonough}, \citenamefont {Hill}, \citenamefont {Simonovi\'c}, \citenamefont
  {Toomey}, \citenamefont {Alexander},\ and\ \citenamefont
  {Zaldarriaga}}]{Ivanov:2020ril}%
  \BibitemOpen
  \bibfield  {author} {\bibinfo {author} {\bibfnamefont {M.~M.}\ \bibnamefont
  {Ivanov}}, \bibinfo {author} {\bibfnamefont {E.}~\bibnamefont {McDonough}},
  \bibinfo {author} {\bibfnamefont {J.~C.}\ \bibnamefont {Hill}}, \bibinfo
  {author} {\bibfnamefont {M.}~\bibnamefont {Simonovi\'c}}, \bibinfo {author}
  {\bibfnamefont {M.~W.}\ \bibnamefont {Toomey}}, \bibinfo {author}
  {\bibfnamefont {S.}~\bibnamefont {Alexander}}, \ and\ \bibinfo {author}
  {\bibfnamefont {M.}~\bibnamefont {Zaldarriaga}},\ }\href {\doibase
  10.1103/PhysRevD.102.103502} {\bibfield  {journal} {\bibinfo  {journal}
  {Phys. Rev. D}\ }\textbf {\bibinfo {volume} {102}},\ \bibinfo {pages}
  {103502} (\bibinfo {year} {2020}{\natexlab{b}})},\ \Eprint
  {http://arxiv.org/abs/2006.11235} {arXiv:2006.11235 [astro-ph.CO]}
  \BibitemShut {NoStop}%
\bibitem [{\citenamefont {D'Amico}\ \emph {et~al.}(2021)\citenamefont
  {D'Amico}, \citenamefont {Senatore}, \citenamefont {Zhang},\ and\
  \citenamefont {Zheng}}]{DAmico:2020ods}%
  \BibitemOpen
  \bibfield  {author} {\bibinfo {author} {\bibfnamefont {G.}~\bibnamefont
  {D'Amico}}, \bibinfo {author} {\bibfnamefont {L.}~\bibnamefont {Senatore}},
  \bibinfo {author} {\bibfnamefont {P.}~\bibnamefont {Zhang}}, \ and\ \bibinfo
  {author} {\bibfnamefont {H.}~\bibnamefont {Zheng}},\ }\href {\doibase
  10.1088/1475-7516/2021/05/072} {\bibfield  {journal} {\bibinfo  {journal}
  {JCAP}\ }\textbf {\bibinfo {volume} {05}},\ \bibinfo {pages} {072} (\bibinfo
  {year} {2021})},\ \Eprint {http://arxiv.org/abs/2006.12420} {arXiv:2006.12420
  [astro-ph.CO]} \BibitemShut {NoStop}%
\bibitem [{\citenamefont {Jedamzik}\ \emph {et~al.}(2021)\citenamefont
  {Jedamzik}, \citenamefont {Pogosian},\ and\ \citenamefont
  {Zhao}}]{Jedamzik:2020zmd}%
  \BibitemOpen
  \bibfield  {author} {\bibinfo {author} {\bibfnamefont {K.}~\bibnamefont
  {Jedamzik}}, \bibinfo {author} {\bibfnamefont {L.}~\bibnamefont {Pogosian}},
  \ and\ \bibinfo {author} {\bibfnamefont {G.-B.}\ \bibnamefont {Zhao}},\
  }\href {\doibase 10.1038/s42005-021-00628-x} {\bibfield  {journal} {\bibinfo
  {journal} {Commun. in Phys.}\ }\textbf {\bibinfo {volume} {4}},\ \bibinfo
  {pages} {123} (\bibinfo {year} {2021})},\ \Eprint
  {http://arxiv.org/abs/2010.04158} {arXiv:2010.04158 [astro-ph.CO]}
  \BibitemShut {NoStop}%
\bibitem [{\citenamefont {Smith}\ \emph {et~al.}(2021)\citenamefont {Smith},
  \citenamefont {Poulin}, \citenamefont {Bernal}, \citenamefont {Boddy},
  \citenamefont {Kamionkowski},\ and\ \citenamefont {Murgia}}]{Smith:2020rxx}%
  \BibitemOpen
  \bibfield  {author} {\bibinfo {author} {\bibfnamefont {T.~L.}\ \bibnamefont
  {Smith}}, \bibinfo {author} {\bibfnamefont {V.}~\bibnamefont {Poulin}},
  \bibinfo {author} {\bibfnamefont {J.~L.}\ \bibnamefont {Bernal}}, \bibinfo
  {author} {\bibfnamefont {K.~K.}\ \bibnamefont {Boddy}}, \bibinfo {author}
  {\bibfnamefont {M.}~\bibnamefont {Kamionkowski}}, \ and\ \bibinfo {author}
  {\bibfnamefont {R.}~\bibnamefont {Murgia}},\ }\href {\doibase
  10.1103/PhysRevD.103.123542} {\bibfield  {journal} {\bibinfo  {journal}
  {Phys. Rev. D}\ }\textbf {\bibinfo {volume} {103}},\ \bibinfo {pages}
  {123542} (\bibinfo {year} {2021})},\ \Eprint
  {http://arxiv.org/abs/2009.10740} {arXiv:2009.10740 [astro-ph.CO]}
  \BibitemShut {NoStop}%
\bibitem [{\citenamefont {Murgia}\ \emph {et~al.}(2021)\citenamefont {Murgia},
  \citenamefont {Abell\'an},\ and\ \citenamefont {Poulin}}]{Murgia:2020ryi}%
  \BibitemOpen
  \bibfield  {author} {\bibinfo {author} {\bibfnamefont {R.}~\bibnamefont
  {Murgia}}, \bibinfo {author} {\bibfnamefont {G.~F.}\ \bibnamefont
  {Abell\'an}}, \ and\ \bibinfo {author} {\bibfnamefont {V.}~\bibnamefont
  {Poulin}},\ }\href {\doibase 10.1103/PhysRevD.103.063502} {\bibfield
  {journal} {\bibinfo  {journal} {Phys. Rev. D}\ }\textbf {\bibinfo {volume}
  {103}},\ \bibinfo {pages} {063502} (\bibinfo {year} {2021})},\ \Eprint
  {http://arxiv.org/abs/2009.10733} {arXiv:2009.10733 [astro-ph.CO]}
  \BibitemShut {NoStop}%
\bibitem [{\citenamefont {Simon}\ \emph {et~al.}(2022)\citenamefont {Simon},
  \citenamefont {Zhang}, \citenamefont {Poulin},\ and\ \citenamefont
  {Smith}}]{Simon:2022adh}%
  \BibitemOpen
  \bibfield  {author} {\bibinfo {author} {\bibfnamefont {T.}~\bibnamefont
  {Simon}}, \bibinfo {author} {\bibfnamefont {P.}~\bibnamefont {Zhang}},
  \bibinfo {author} {\bibfnamefont {V.}~\bibnamefont {Poulin}}, \ and\ \bibinfo
  {author} {\bibfnamefont {T.~L.}\ \bibnamefont {Smith}},\ }\href@noop {} {\
  (\bibinfo {year} {2022})},\ \Eprint {http://arxiv.org/abs/2208.05930}
  {arXiv:2208.05930 [astro-ph.CO]} \BibitemShut {NoStop}%
\bibitem [{\citenamefont {Herold}\ \emph {et~al.}(2022)\citenamefont {Herold},
  \citenamefont {Ferreira},\ and\ \citenamefont {Komatsu}}]{Herold:2021ksg}%
  \BibitemOpen
  \bibfield  {author} {\bibinfo {author} {\bibfnamefont {L.}~\bibnamefont
  {Herold}}, \bibinfo {author} {\bibfnamefont {E.~G.~M.}\ \bibnamefont
  {Ferreira}}, \ and\ \bibinfo {author} {\bibfnamefont {E.}~\bibnamefont
  {Komatsu}},\ }\href {\doibase 10.3847/2041-8213/ac63a3} {\bibfield  {journal}
  {\bibinfo  {journal} {Astrophys. J. Lett.}\ }\textbf {\bibinfo {volume}
  {929}},\ \bibinfo {pages} {L16} (\bibinfo {year} {2022})},\ \Eprint
  {http://arxiv.org/abs/2112.12140} {arXiv:2112.12140 [astro-ph.CO]}
  \BibitemShut {NoStop}%
\bibitem [{\citenamefont {Vagnozzi}(2021)}]{Vagnozzi:2021gjh}%
  \BibitemOpen
  \bibfield  {author} {\bibinfo {author} {\bibfnamefont {S.}~\bibnamefont
  {Vagnozzi}},\ }\href {\doibase 10.1103/PhysRevD.104.063524} {\bibfield
  {journal} {\bibinfo  {journal} {Phys. Rev. D}\ }\textbf {\bibinfo {volume}
  {104}},\ \bibinfo {pages} {063524} (\bibinfo {year} {2021})},\ \Eprint
  {http://arxiv.org/abs/2105.10425} {arXiv:2105.10425 [astro-ph.CO]}
  \BibitemShut {NoStop}%
\bibitem [{\citenamefont {Chabanier}\ \emph
  {et~al.}(2019{\natexlab{a}})\citenamefont {Chabanier} \emph
  {et~al.}}]{Chabanier:2018rga}%
  \BibitemOpen
  \bibfield  {author} {\bibinfo {author} {\bibfnamefont {S.}~\bibnamefont
  {Chabanier}} \emph {et~al.},\ }\href {\doibase 10.1088/1475-7516/2019/07/017}
  {\bibfield  {journal} {\bibinfo  {journal} {JCAP}\ }\textbf {\bibinfo
  {volume} {07}},\ \bibinfo {pages} {017} (\bibinfo {year}
  {2019}{\natexlab{a}})},\ \Eprint {http://arxiv.org/abs/1812.03554}
  {arXiv:1812.03554 [astro-ph.CO]} \BibitemShut {NoStop}%
\bibitem [{\citenamefont {Palanque-Delabrouille}\ \emph
  {et~al.}(2020)\citenamefont {Palanque-Delabrouille}, \citenamefont {Y\`eche},
  \citenamefont {Sch\"oneberg}, \citenamefont {Lesgourgues}, \citenamefont
  {Walther}, \citenamefont {Chabanier},\ and\ \citenamefont
  {Armengaud}}]{Palanque-Delabrouille:2019iyz}%
  \BibitemOpen
  \bibfield  {author} {\bibinfo {author} {\bibfnamefont {N.}~\bibnamefont
  {Palanque-Delabrouille}}, \bibinfo {author} {\bibfnamefont {C.}~\bibnamefont
  {Y\`eche}}, \bibinfo {author} {\bibfnamefont {N.}~\bibnamefont
  {Sch\"oneberg}}, \bibinfo {author} {\bibfnamefont {J.}~\bibnamefont
  {Lesgourgues}}, \bibinfo {author} {\bibfnamefont {M.}~\bibnamefont
  {Walther}}, \bibinfo {author} {\bibfnamefont {S.}~\bibnamefont {Chabanier}},
  \ and\ \bibinfo {author} {\bibfnamefont {E.}~\bibnamefont {Armengaud}},\
  }\href {\doibase 10.1088/1475-7516/2020/04/038} {\bibfield  {journal}
  {\bibinfo  {journal} {JCAP}\ }\textbf {\bibinfo {volume} {04}},\ \bibinfo
  {pages} {038} (\bibinfo {year} {2020})},\ \Eprint
  {http://arxiv.org/abs/1911.09073} {arXiv:1911.09073 [astro-ph.CO]}
  \BibitemShut {NoStop}%
\bibitem [{\citenamefont {Esposito}\ \emph {et~al.}(2022)\citenamefont
  {Esposito}, \citenamefont {Ir\v{s}i\v{c}}, \citenamefont {Costanzi},
  \citenamefont {Borgani}, \citenamefont {Saro},\ and\ \citenamefont
  {Viel}}]{Esposito:2022plo}%
  \BibitemOpen
  \bibfield  {author} {\bibinfo {author} {\bibfnamefont {M.}~\bibnamefont
  {Esposito}}, \bibinfo {author} {\bibfnamefont {V.}~\bibnamefont
  {Ir\v{s}i\v{c}}}, \bibinfo {author} {\bibfnamefont {M.}~\bibnamefont
  {Costanzi}}, \bibinfo {author} {\bibfnamefont {S.}~\bibnamefont {Borgani}},
  \bibinfo {author} {\bibfnamefont {A.}~\bibnamefont {Saro}}, \ and\ \bibinfo
  {author} {\bibfnamefont {M.}~\bibnamefont {Viel}},\ }\href {\doibase
  10.1093/mnras/stac1825} {\bibfield  {journal} {\bibinfo  {journal} {Mon. Not.
  Roy. Astron. Soc.}\ }\textbf {\bibinfo {volume} {515}},\ \bibinfo {pages}
  {857} (\bibinfo {year} {2022})},\ \Eprint {http://arxiv.org/abs/2202.00974}
  {arXiv:2202.00974 [astro-ph.CO]} \BibitemShut {NoStop}%
\bibitem [{\citenamefont {Smith}\ \emph {et~al.}(2020)\citenamefont {Smith},
  \citenamefont {Poulin},\ and\ \citenamefont {Amin}}]{Smith:2019ihp}%
  \BibitemOpen
  \bibfield  {author} {\bibinfo {author} {\bibfnamefont {T.~L.}\ \bibnamefont
  {Smith}}, \bibinfo {author} {\bibfnamefont {V.}~\bibnamefont {Poulin}}, \
  and\ \bibinfo {author} {\bibfnamefont {M.~A.}\ \bibnamefont {Amin}},\ }\href
  {\doibase 10.1103/PhysRevD.101.063523} {\bibfield  {journal} {\bibinfo
  {journal} {Phys. Rev. D}\ }\textbf {\bibinfo {volume} {101}},\ \bibinfo
  {pages} {063523} (\bibinfo {year} {2020})},\ \Eprint
  {http://arxiv.org/abs/1908.06995} {arXiv:1908.06995 [astro-ph.CO]}
  \BibitemShut {NoStop}%
\bibitem [{\citenamefont {Lesgourgues}(2011)}]{Lesgourgues:2011re}%
  \BibitemOpen
  \bibfield  {author} {\bibinfo {author} {\bibfnamefont {J.}~\bibnamefont
  {Lesgourgues}},\ }\href@noop {} {\  (\bibinfo {year} {2011})},\ \Eprint
  {http://arxiv.org/abs/1104.2932} {arXiv:1104.2932 [astro-ph.IM]} \BibitemShut
  {NoStop}%
\bibitem [{\citenamefont {Blas}\ \emph {et~al.}(2011)\citenamefont {Blas},
  \citenamefont {Lesgourgues},\ and\ \citenamefont {Tram}}]{Blas:2011rf}%
  \BibitemOpen
  \bibfield  {author} {\bibinfo {author} {\bibfnamefont {D.}~\bibnamefont
  {Blas}}, \bibinfo {author} {\bibfnamefont {J.}~\bibnamefont {Lesgourgues}}, \
  and\ \bibinfo {author} {\bibfnamefont {T.}~\bibnamefont {Tram}},\ }\href
  {\doibase 10.1088/1475-7516/2011/07/034} {\bibfield  {journal} {\bibinfo
  {journal} {JCAP}\ }\textbf {\bibinfo {volume} {07}},\ \bibinfo {pages} {034}
  (\bibinfo {year} {2011})},\ \Eprint {http://arxiv.org/abs/1104.2933}
  {arXiv:1104.2933 [astro-ph.CO]} \BibitemShut {NoStop}%
\bibitem [{\citenamefont {Aghanim}\ \emph
  {et~al.}(2020{\natexlab{b}})\citenamefont {Aghanim} \emph
  {et~al.}}]{Aghanim:2019ame}%
  \BibitemOpen
  \bibfield  {author} {\bibinfo {author} {\bibfnamefont {N.}~\bibnamefont
  {Aghanim}} \emph {et~al.} (\bibinfo {collaboration} {Planck}),\ }\href
  {\doibase 10.1051/0004-6361/201936386} {\bibfield  {journal} {\bibinfo
  {journal} {Astron. Astrophys.}\ }\textbf {\bibinfo {volume} {641}},\ \bibinfo
  {pages} {A5} (\bibinfo {year} {2020}{\natexlab{b}})},\ \Eprint
  {http://arxiv.org/abs/1907.12875} {arXiv:1907.12875 [astro-ph.CO]}
  \BibitemShut {NoStop}%
\bibitem [{\citenamefont {Aghanim}\ \emph
  {et~al.}(2020{\natexlab{c}})\citenamefont {Aghanim} \emph
  {et~al.}}]{Aghanim:2018oex}%
  \BibitemOpen
  \bibfield  {author} {\bibinfo {author} {\bibfnamefont {N.}~\bibnamefont
  {Aghanim}} \emph {et~al.} (\bibinfo {collaboration} {Planck}),\ }\href
  {\doibase 10.1051/0004-6361/201833886} {\bibfield  {journal} {\bibinfo
  {journal} {Astron. Astrophys.}\ }\textbf {\bibinfo {volume} {641}},\ \bibinfo
  {pages} {A8} (\bibinfo {year} {2020}{\natexlab{c}})},\ \Eprint
  {http://arxiv.org/abs/1807.06210} {arXiv:1807.06210 [astro-ph.CO]}
  \BibitemShut {NoStop}%
\bibitem [{\citenamefont {Alam}\ \emph {et~al.}(2017)\citenamefont {Alam} \emph
  {et~al.}}]{Alam:2016hwk}%
  \BibitemOpen
  \bibfield  {author} {\bibinfo {author} {\bibfnamefont {S.}~\bibnamefont
  {Alam}} \emph {et~al.} (\bibinfo {collaboration} {BOSS}),\ }\href {\doibase
  10.1093/mnras/stx721} {\bibfield  {journal} {\bibinfo  {journal} {Mon. Not.
  Roy. Astron. Soc.}\ }\textbf {\bibinfo {volume} {470}},\ \bibinfo {pages}
  {2617} (\bibinfo {year} {2017})},\ \Eprint {http://arxiv.org/abs/1607.03155}
  {arXiv:1607.03155 [astro-ph.CO]} \BibitemShut {NoStop}%
\bibitem [{\citenamefont {Ross}\ \emph {et~al.}(2015)\citenamefont {Ross},
  \citenamefont {Samushia}, \citenamefont {Howlett}, \citenamefont {Percival},
  \citenamefont {Burden},\ and\ \citenamefont {Manera}}]{Ross:2014qpa}%
  \BibitemOpen
  \bibfield  {author} {\bibinfo {author} {\bibfnamefont {A.~J.}\ \bibnamefont
  {Ross}}, \bibinfo {author} {\bibfnamefont {L.}~\bibnamefont {Samushia}},
  \bibinfo {author} {\bibfnamefont {C.}~\bibnamefont {Howlett}}, \bibinfo
  {author} {\bibfnamefont {W.~J.}\ \bibnamefont {Percival}}, \bibinfo {author}
  {\bibfnamefont {A.}~\bibnamefont {Burden}}, \ and\ \bibinfo {author}
  {\bibfnamefont {M.}~\bibnamefont {Manera}},\ }\href {\doibase
  10.1093/mnras/stv154} {\bibfield  {journal} {\bibinfo  {journal} {Mon. Not.
  Roy. Astron. Soc.}\ }\textbf {\bibinfo {volume} {449}},\ \bibinfo {pages}
  {835} (\bibinfo {year} {2015})},\ \Eprint {http://arxiv.org/abs/1409.3242}
  {arXiv:1409.3242 [astro-ph.CO]} \BibitemShut {NoStop}%
\bibitem [{\citenamefont {Beutler}\ \emph {et~al.}(2011)\citenamefont
  {Beutler}, \citenamefont {Blake}, \citenamefont {Colless}, \citenamefont
  {Jones}, \citenamefont {Staveley-Smith}, \citenamefont {Campbell},
  \citenamefont {Parker}, \citenamefont {Saunders},\ and\ \citenamefont
  {Watson}}]{Beutler:2011hx}%
  \BibitemOpen
  \bibfield  {author} {\bibinfo {author} {\bibfnamefont {F.}~\bibnamefont
  {Beutler}}, \bibinfo {author} {\bibfnamefont {C.}~\bibnamefont {Blake}},
  \bibinfo {author} {\bibfnamefont {M.}~\bibnamefont {Colless}}, \bibinfo
  {author} {\bibfnamefont {D.}~\bibnamefont {Jones}}, \bibinfo {author}
  {\bibfnamefont {L.}~\bibnamefont {Staveley-Smith}}, \bibinfo {author}
  {\bibfnamefont {L.}~\bibnamefont {Campbell}}, \bibinfo {author}
  {\bibfnamefont {Q.}~\bibnamefont {Parker}}, \bibinfo {author} {\bibfnamefont
  {W.}~\bibnamefont {Saunders}}, \ and\ \bibinfo {author} {\bibfnamefont
  {F.}~\bibnamefont {Watson}},\ }\href {\doibase
  10.1111/j.1365-2966.2011.19250.x} {\bibfield  {journal} {\bibinfo  {journal}
  {Mon. Not. Roy. Astron. Soc.}\ }\textbf {\bibinfo {volume} {416}},\ \bibinfo
  {pages} {3017} (\bibinfo {year} {2011})},\ \Eprint
  {http://arxiv.org/abs/1106.3366} {arXiv:1106.3366 [astro-ph.CO]} \BibitemShut
  {NoStop}%
\bibitem [{\citenamefont {McDonald}\ \emph {et~al.}(2005)\citenamefont
  {McDonald} \emph {et~al.}}]{SDSS:2004aee}%
  \BibitemOpen
  \bibfield  {author} {\bibinfo {author} {\bibfnamefont {P.}~\bibnamefont
  {McDonald}} \emph {et~al.} (\bibinfo {collaboration} {SDSS}),\ }\href
  {\doibase 10.1086/497563} {\bibfield  {journal} {\bibinfo  {journal}
  {Astrophys. J.}\ }\textbf {\bibinfo {volume} {635}},\ \bibinfo {pages} {761}
  (\bibinfo {year} {2005})},\ \Eprint {http://arxiv.org/abs/astro-ph/0407377}
  {arXiv:astro-ph/0407377} \BibitemShut {NoStop}%
\bibitem [{\citenamefont {Pedersen}\ \emph {et~al.}(2022)\citenamefont
  {Pedersen}, \citenamefont {Font-Ribera},\ and\ \citenamefont
  {Gnedin}}]{Pedersen:2022anu}%
  \BibitemOpen
  \bibfield  {author} {\bibinfo {author} {\bibfnamefont {C.}~\bibnamefont
  {Pedersen}}, \bibinfo {author} {\bibfnamefont {A.}~\bibnamefont
  {Font-Ribera}}, \ and\ \bibinfo {author} {\bibfnamefont {N.~Y.}\ \bibnamefont
  {Gnedin}},\ }\href@noop {} {\  (\bibinfo {year} {2022})},\ \Eprint
  {http://arxiv.org/abs/2209.09895} {arXiv:2209.09895 [astro-ph.CO]}
  \BibitemShut {NoStop}%
\bibitem [{\citenamefont {Seljak}\ \emph {et~al.}(2006)\citenamefont {Seljak},
  \citenamefont {Slosar},\ and\ \citenamefont {McDonald}}]{Seljak:2006bg}%
  \BibitemOpen
  \bibfield  {author} {\bibinfo {author} {\bibfnamefont {U.}~\bibnamefont
  {Seljak}}, \bibinfo {author} {\bibfnamefont {A.}~\bibnamefont {Slosar}}, \
  and\ \bibinfo {author} {\bibfnamefont {P.}~\bibnamefont {McDonald}},\ }\href
  {\doibase 10.1088/1475-7516/2006/10/014} {\bibfield  {journal} {\bibinfo
  {journal} {JCAP}\ }\textbf {\bibinfo {volume} {10}},\ \bibinfo {pages} {014}
  (\bibinfo {year} {2006})},\ \Eprint {http://arxiv.org/abs/astro-ph/0604335}
  {arXiv:astro-ph/0604335} \BibitemShut {NoStop}%
\bibitem [{\citenamefont {Ir\v{s}i\v{c}}\ \emph
  {et~al.}(2017{\natexlab{a}})\citenamefont {Ir\v{s}i\v{c}} \emph
  {et~al.}}]{Irsic:2017sop}%
  \BibitemOpen
  \bibfield  {author} {\bibinfo {author} {\bibfnamefont {V.}~\bibnamefont
  {Ir\v{s}i\v{c}}} \emph {et~al.},\ }\href {\doibase 10.1093/mnras/stw3372}
  {\bibfield  {journal} {\bibinfo  {journal} {Mon. Not. Roy. Astron. Soc.}\
  }\textbf {\bibinfo {volume} {466}},\ \bibinfo {pages} {4332} (\bibinfo {year}
  {2017}{\natexlab{a}})},\ \Eprint {http://arxiv.org/abs/1702.01761}
  {arXiv:1702.01761 [astro-ph.CO]} \BibitemShut {NoStop}%
\bibitem [{\citenamefont {Viel}\ \emph {et~al.}(2013)\citenamefont {Viel},
  \citenamefont {Becker}, \citenamefont {Bolton},\ and\ \citenamefont
  {Haehnelt}}]{Viel:2013fqw}%
  \BibitemOpen
  \bibfield  {author} {\bibinfo {author} {\bibfnamefont {M.}~\bibnamefont
  {Viel}}, \bibinfo {author} {\bibfnamefont {G.~D.}\ \bibnamefont {Becker}},
  \bibinfo {author} {\bibfnamefont {J.~S.}\ \bibnamefont {Bolton}}, \ and\
  \bibinfo {author} {\bibfnamefont {M.~G.}\ \bibnamefont {Haehnelt}},\ }\href
  {\doibase 10.1103/PhysRevD.88.043502} {\bibfield  {journal} {\bibinfo
  {journal} {Phys. Rev. D}\ }\textbf {\bibinfo {volume} {88}},\ \bibinfo
  {pages} {043502} (\bibinfo {year} {2013})},\ \Eprint
  {http://arxiv.org/abs/1306.2314} {arXiv:1306.2314 [astro-ph.CO]} \BibitemShut
  {NoStop}%
\bibitem [{\citenamefont {Torrado}\ and\ \citenamefont
  {Lewis}(2021)}]{Torrado:2020dgo}%
  \BibitemOpen
  \bibfield  {author} {\bibinfo {author} {\bibfnamefont {J.}~\bibnamefont
  {Torrado}}\ and\ \bibinfo {author} {\bibfnamefont {A.}~\bibnamefont
  {Lewis}},\ }\href {\doibase 10.1088/1475-7516/2021/05/057} {\bibfield
  {journal} {\bibinfo  {journal} {JCAP}\ }\textbf {\bibinfo {volume} {05}},\
  \bibinfo {pages} {057} (\bibinfo {year} {2021})},\ \Eprint
  {http://arxiv.org/abs/2005.05290} {arXiv:2005.05290 [astro-ph.IM]}
  \BibitemShut {NoStop}%
\bibitem [{\citenamefont {Gelman}\ and\ \citenamefont
  {Rubin}(1992)}]{gelman1992}%
  \BibitemOpen
  \bibfield  {author} {\bibinfo {author} {\bibfnamefont {A.}~\bibnamefont
  {Gelman}}\ and\ \bibinfo {author} {\bibfnamefont {D.~B.}\ \bibnamefont
  {Rubin}},\ }\href {\doibase 10.1214/ss/1177011136} {\bibfield  {journal}
  {\bibinfo  {journal} {Statist. Sci.}\ }\textbf {\bibinfo {volume} {7}},\
  \bibinfo {pages} {457} (\bibinfo {year} {1992})}\BibitemShut {NoStop}%
\bibitem [{\citenamefont {Lewis}(2019)}]{Lewis:2019xzd}%
  \BibitemOpen
  \bibfield  {author} {\bibinfo {author} {\bibfnamefont {A.}~\bibnamefont
  {Lewis}},\ }\href@noop {} {\  (\bibinfo {year} {2019})},\ \Eprint
  {http://arxiv.org/abs/1910.13970} {arXiv:1910.13970 [astro-ph.IM]}
  \BibitemShut {NoStop}%
\bibitem [{\citenamefont {Sch\"oneberg}\ \emph
  {et~al.}(2022{\natexlab{b}})\citenamefont {Sch\"oneberg}, \citenamefont
  {Franco~Abell\'an}, \citenamefont {P\'erez~S\'anchez}, \citenamefont {Witte},
  \citenamefont {Poulin},\ and\ \citenamefont
  {Lesgourgues}}]{Schoneberg:2021qvd}%
  \BibitemOpen
  \bibfield  {author} {\bibinfo {author} {\bibfnamefont {N.}~\bibnamefont
  {Sch\"oneberg}}, \bibinfo {author} {\bibfnamefont {G.}~\bibnamefont
  {Franco~Abell\'an}}, \bibinfo {author} {\bibfnamefont {A.}~\bibnamefont
  {P\'erez~S\'anchez}}, \bibinfo {author} {\bibfnamefont {S.~J.}\ \bibnamefont
  {Witte}}, \bibinfo {author} {\bibfnamefont {V.}~\bibnamefont {Poulin}}, \
  and\ \bibinfo {author} {\bibfnamefont {J.}~\bibnamefont {Lesgourgues}},\
  }\href {\doibase 10.1016/j.physrep.2022.07.001} {\bibfield  {journal}
  {\bibinfo  {journal} {Phys. Rept.}\ }\textbf {\bibinfo {volume} {984}},\
  \bibinfo {pages} {1} (\bibinfo {year} {2022}{\natexlab{b}})},\ \Eprint
  {http://arxiv.org/abs/2107.10291} {arXiv:2107.10291 [astro-ph.CO]}
  \BibitemShut {NoStop}%
\bibitem [{\citenamefont {Hill}\ \emph {et~al.}(2022)\citenamefont {Hill} \emph
  {et~al.}}]{Hill:2021yec}%
  \BibitemOpen
  \bibfield  {author} {\bibinfo {author} {\bibfnamefont {J.~C.}\ \bibnamefont
  {Hill}} \emph {et~al.},\ }\href {\doibase 10.1103/PhysRevD.105.123536}
  {\bibfield  {journal} {\bibinfo  {journal} {Phys. Rev. D}\ }\textbf {\bibinfo
  {volume} {105}},\ \bibinfo {pages} {123536} (\bibinfo {year} {2022})},\
  \Eprint {http://arxiv.org/abs/2109.04451} {arXiv:2109.04451 [astro-ph.CO]}
  \BibitemShut {NoStop}%
\bibitem [{\citenamefont {La~Posta}\ \emph {et~al.}(2022)\citenamefont
  {La~Posta}, \citenamefont {Louis}, \citenamefont {Garrido},\ and\
  \citenamefont {Hill}}]{LaPosta:2021pgm}%
  \BibitemOpen
  \bibfield  {author} {\bibinfo {author} {\bibfnamefont {A.}~\bibnamefont
  {La~Posta}}, \bibinfo {author} {\bibfnamefont {T.}~\bibnamefont {Louis}},
  \bibinfo {author} {\bibfnamefont {X.}~\bibnamefont {Garrido}}, \ and\
  \bibinfo {author} {\bibfnamefont {J.~C.}\ \bibnamefont {Hill}},\ }\href
  {\doibase 10.1103/PhysRevD.105.083519} {\bibfield  {journal} {\bibinfo
  {journal} {Phys. Rev. D}\ }\textbf {\bibinfo {volume} {105}},\ \bibinfo
  {pages} {083519} (\bibinfo {year} {2022})},\ \Eprint
  {http://arxiv.org/abs/2112.10754} {arXiv:2112.10754 [astro-ph.CO]}
  \BibitemShut {NoStop}%
\bibitem [{\citenamefont {Herold}\ and\ \citenamefont
  {Ferreira}(2022)}]{Herold:2022iib}%
  \BibitemOpen
  \bibfield  {author} {\bibinfo {author} {\bibfnamefont {L.}~\bibnamefont
  {Herold}}\ and\ \bibinfo {author} {\bibfnamefont {E.~G.~M.}\ \bibnamefont
  {Ferreira}},\ }\href@noop {} {\  (\bibinfo {year} {2022})},\ \Eprint
  {http://arxiv.org/abs/2210.16296} {arXiv:2210.16296 [astro-ph.CO]}
  \BibitemShut {NoStop}%
\bibitem [{\citenamefont {G\'omez-Valent}(2022)}]{Gomez-Valent:2022hkb}%
  \BibitemOpen
  \bibfield  {author} {\bibinfo {author} {\bibfnamefont {A.}~\bibnamefont
  {G\'omez-Valent}},\ }\href {\doibase 10.1103/PhysRevD.106.063506} {\bibfield
  {journal} {\bibinfo  {journal} {Phys. Rev. D}\ }\textbf {\bibinfo {volume}
  {106}},\ \bibinfo {pages} {063506} (\bibinfo {year} {2022})},\ \Eprint
  {http://arxiv.org/abs/2203.16285} {arXiv:2203.16285 [astro-ph.CO]}
  \BibitemShut {NoStop}%
\bibitem [{\citenamefont {Villasenor}\ \emph {et~al.}(2022)\citenamefont
  {Villasenor}, \citenamefont {Robertson}, \citenamefont {Madau},\ and\
  \citenamefont {Schneider}}]{Villasenor:2022aiy}%
  \BibitemOpen
  \bibfield  {author} {\bibinfo {author} {\bibfnamefont {B.}~\bibnamefont
  {Villasenor}}, \bibinfo {author} {\bibfnamefont {B.}~\bibnamefont
  {Robertson}}, \bibinfo {author} {\bibfnamefont {P.}~\bibnamefont {Madau}}, \
  and\ \bibinfo {author} {\bibfnamefont {E.}~\bibnamefont {Schneider}},\
  }\href@noop {} {\  (\bibinfo {year} {2022})},\ \Eprint
  {http://arxiv.org/abs/2209.14220} {arXiv:2209.14220 [astro-ph.CO]}
  \BibitemShut {NoStop}%
\bibitem [{\citenamefont {Fernandez}\ \emph {et~al.}(2023)\citenamefont
  {Fernandez}, \citenamefont {Bird},\ and\ \citenamefont
  {Ho}}]{Fernandez:2023grg}%
  \BibitemOpen
  \bibfield  {author} {\bibinfo {author} {\bibfnamefont {M.~A.}\ \bibnamefont
  {Fernandez}}, \bibinfo {author} {\bibfnamefont {S.}~\bibnamefont {Bird}}, \
  and\ \bibinfo {author} {\bibfnamefont {M.-F.}\ \bibnamefont {Ho}},\
  }\href@noop {} {\  (\bibinfo {year} {2023})},\ \Eprint
  {http://arxiv.org/abs/2309.03943} {arXiv:2309.03943 [astro-ph.CO]}
  \BibitemShut {NoStop}%
\bibitem [{\citenamefont {McDonough}\ \emph {et~al.}(2022)\citenamefont
  {McDonough}, \citenamefont {Lin}, \citenamefont {Hill}, \citenamefont {Hu},\
  and\ \citenamefont {Zhou}}]{McDonough:2021pdg}%
  \BibitemOpen
  \bibfield  {author} {\bibinfo {author} {\bibfnamefont {E.}~\bibnamefont
  {McDonough}}, \bibinfo {author} {\bibfnamefont {M.-X.}\ \bibnamefont {Lin}},
  \bibinfo {author} {\bibfnamefont {J.~C.}\ \bibnamefont {Hill}}, \bibinfo
  {author} {\bibfnamefont {W.}~\bibnamefont {Hu}}, \ and\ \bibinfo {author}
  {\bibfnamefont {S.}~\bibnamefont {Zhou}},\ }\href {\doibase
  10.1103/PhysRevD.106.043525} {\bibfield  {journal} {\bibinfo  {journal}
  {Phys. Rev. D}\ }\textbf {\bibinfo {volume} {106}},\ \bibinfo {pages}
  {043525} (\bibinfo {year} {2022})},\ \Eprint
  {http://arxiv.org/abs/2112.09128} {arXiv:2112.09128 [astro-ph.CO]}
  \BibitemShut {NoStop}%
\bibitem [{\citenamefont {Karwal}\ \emph {et~al.}(2022)\citenamefont {Karwal},
  \citenamefont {Raveri}, \citenamefont {Jain}, \citenamefont {Khoury},\ and\
  \citenamefont {Trodden}}]{Karwal:2021vpk}%
  \BibitemOpen
  \bibfield  {author} {\bibinfo {author} {\bibfnamefont {T.}~\bibnamefont
  {Karwal}}, \bibinfo {author} {\bibfnamefont {M.}~\bibnamefont {Raveri}},
  \bibinfo {author} {\bibfnamefont {B.}~\bibnamefont {Jain}}, \bibinfo {author}
  {\bibfnamefont {J.}~\bibnamefont {Khoury}}, \ and\ \bibinfo {author}
  {\bibfnamefont {M.}~\bibnamefont {Trodden}},\ }\href {\doibase
  10.1103/PhysRevD.105.063535} {\bibfield  {journal} {\bibinfo  {journal}
  {Phys. Rev. D}\ }\textbf {\bibinfo {volume} {105}},\ \bibinfo {pages}
  {063535} (\bibinfo {year} {2022})},\ \Eprint
  {http://arxiv.org/abs/2106.13290} {arXiv:2106.13290 [astro-ph.CO]}
  \BibitemShut {NoStop}%
\bibitem [{\citenamefont {Lin}\ \emph {et~al.}(2022)\citenamefont {Lin},
  \citenamefont {McDonough}, \citenamefont {Hill},\ and\ \citenamefont
  {Hu}}]{Lin:2022phm}%
  \BibitemOpen
  \bibfield  {author} {\bibinfo {author} {\bibfnamefont {M.-X.}\ \bibnamefont
  {Lin}}, \bibinfo {author} {\bibfnamefont {E.}~\bibnamefont {McDonough}},
  \bibinfo {author} {\bibfnamefont {J.~C.}\ \bibnamefont {Hill}}, \ and\
  \bibinfo {author} {\bibfnamefont {W.}~\bibnamefont {Hu}},\ }\href@noop {} {\
  (\bibinfo {year} {2022})},\ \Eprint {http://arxiv.org/abs/2212.08098}
  {arXiv:2212.08098 [astro-ph.CO]} \BibitemShut {NoStop}%
\bibitem [{\citenamefont {Berghaus}\ and\ \citenamefont
  {Karwal}(2022)}]{Berghaus:2022cwf}%
  \BibitemOpen
  \bibfield  {author} {\bibinfo {author} {\bibfnamefont {K.~V.}\ \bibnamefont
  {Berghaus}}\ and\ \bibinfo {author} {\bibfnamefont {T.}~\bibnamefont
  {Karwal}},\ }\href@noop {} {\  (\bibinfo {year} {2022})},\ \Eprint
  {http://arxiv.org/abs/2204.09133} {arXiv:2204.09133 [astro-ph.CO]}
  \BibitemShut {NoStop}%
\bibitem [{\citenamefont {Abbott}\ \emph {et~al.}(2022)\citenamefont {Abbott}
  \emph {et~al.}}]{DES_Y3}%
  \BibitemOpen
  \bibfield  {author} {\bibinfo {author} {\bibfnamefont {T.~M.~C.}\
  \bibnamefont {Abbott}} \emph {et~al.} (\bibinfo {collaboration} {DES}),\
  }\href {\doibase 10.1103/PhysRevD.105.023520} {\bibfield  {journal} {\bibinfo
   {journal} {Phys. Rev. D}\ }\textbf {\bibinfo {volume} {105}},\ \bibinfo
  {pages} {023520} (\bibinfo {year} {2022})},\ \Eprint
  {http://arxiv.org/abs/2105.13549} {arXiv:2105.13549 [astro-ph.CO]}
  \BibitemShut {NoStop}%
\bibitem [{\citenamefont {D'Amico}\ \emph {et~al.}(2019)\citenamefont
  {D'Amico}, \citenamefont {Gleyzes}, \citenamefont {Kokron}, \citenamefont
  {Markovic}, \citenamefont {Senatore}, \citenamefont {Zhang}, \citenamefont
  {Beutler},\ and\ \citenamefont {Gil-Marín}}]{DAmico:2019fhj}%
  \BibitemOpen
  \bibfield  {author} {\bibinfo {author} {\bibfnamefont {G.}~\bibnamefont
  {D'Amico}}, \bibinfo {author} {\bibfnamefont {J.}~\bibnamefont {Gleyzes}},
  \bibinfo {author} {\bibfnamefont {N.}~\bibnamefont {Kokron}}, \bibinfo
  {author} {\bibfnamefont {D.}~\bibnamefont {Markovic}}, \bibinfo {author}
  {\bibfnamefont {L.}~\bibnamefont {Senatore}}, \bibinfo {author}
  {\bibfnamefont {P.}~\bibnamefont {Zhang}}, \bibinfo {author} {\bibfnamefont
  {F.}~\bibnamefont {Beutler}}, \ and\ \bibinfo {author} {\bibfnamefont
  {H.}~\bibnamefont {Gil-Marín}},\ }\href@noop {} {\  (\bibinfo {year}
  {2019})},\ \Eprint {http://arxiv.org/abs/1909.05271} {arXiv:1909.05271
  [astro-ph.CO]} \BibitemShut {NoStop}%
%%CITATION = ARXIV:1909.05271;%%
\bibitem [{\citenamefont {Poulin}\ \emph {et~al.}(2021)\citenamefont {Poulin},
  \citenamefont {Smith},\ and\ \citenamefont {Bartlett}}]{Poulin:2021bjr}%
  \BibitemOpen
  \bibfield  {author} {\bibinfo {author} {\bibfnamefont {V.}~\bibnamefont
  {Poulin}}, \bibinfo {author} {\bibfnamefont {T.~L.}\ \bibnamefont {Smith}}, \
  and\ \bibinfo {author} {\bibfnamefont {A.}~\bibnamefont {Bartlett}},\ }\href
  {\doibase 10.1103/PhysRevD.104.123550} {\bibfield  {journal} {\bibinfo
  {journal} {Phys. Rev. D}\ }\textbf {\bibinfo {volume} {104}},\ \bibinfo
  {pages} {123550} (\bibinfo {year} {2021})},\ \Eprint
  {http://arxiv.org/abs/2109.06229} {arXiv:2109.06229 [astro-ph.CO]}
  \BibitemShut {NoStop}%
\bibitem [{\citenamefont {Choi}\ \emph {et~al.}(2020)\citenamefont {Choi} \emph
  {et~al.}}]{ACT:2020frw}%
  \BibitemOpen
  \bibfield  {author} {\bibinfo {author} {\bibfnamefont {S.~K.}\ \bibnamefont
  {Choi}} \emph {et~al.} (\bibinfo {collaboration} {ACT}),\ }\href {\doibase
  10.1088/1475-7516/2020/12/045} {\bibfield  {journal} {\bibinfo  {journal}
  {JCAP}\ }\textbf {\bibinfo {volume} {12}},\ \bibinfo {pages} {045} (\bibinfo
  {year} {2020})},\ \Eprint {http://arxiv.org/abs/2007.07289} {arXiv:2007.07289
  [astro-ph.CO]} \BibitemShut {NoStop}%
\bibitem [{\citenamefont {Aghamousa}\ \emph {et~al.}(2016)\citenamefont
  {Aghamousa} \emph {et~al.}}]{DESI:2016fyo}%
  \BibitemOpen
  \bibfield  {author} {\bibinfo {author} {\bibfnamefont {A.}~\bibnamefont
  {Aghamousa}} \emph {et~al.} (\bibinfo {collaboration} {DESI}),\ }\href@noop
  {} {\  (\bibinfo {year} {2016})},\ \Eprint {http://arxiv.org/abs/1611.00036}
  {arXiv:1611.00036 [astro-ph.IM]} \BibitemShut {NoStop}%
\bibitem [{\citenamefont {Ramirez-Perez}\ \emph {et~al.}(2023)\citenamefont
  {Ramirez-Perez} \emph {et~al.}}]{DESI:2023pir}%
  \BibitemOpen
  \bibfield  {author} {\bibinfo {author} {\bibfnamefont {C.}~\bibnamefont
  {Ramirez-Perez}} \emph {et~al.} (\bibinfo {collaboration} {DESI}),\
  }\href@noop {} {\  (\bibinfo {year} {2023})},\ \Eprint
  {http://arxiv.org/abs/2306.06312} {arXiv:2306.06312 [astro-ph.CO]}
  \BibitemShut {NoStop}%
\bibitem [{\citenamefont {Ravoux}\ \emph {et~al.}(2023)\citenamefont {Ravoux}
  \emph {et~al.}}]{DESI:2023xwh}%
  \BibitemOpen
  \bibfield  {author} {\bibinfo {author} {\bibfnamefont {C.}~\bibnamefont
  {Ravoux}} \emph {et~al.} (\bibinfo {collaboration} {DESI}),\ }\href@noop {}
  {\  (\bibinfo {year} {2023})},\ \Eprint {http://arxiv.org/abs/2306.06311}
  {arXiv:2306.06311 [astro-ph.CO]} \BibitemShut {NoStop}%
\bibitem [{\citenamefont {Kara\c{c}ayl\i{}}\ \emph {et~al.}(2023)\citenamefont
  {Kara\c{c}ayl\i{}} \emph {et~al.}}]{Karacayli:2023afs}%
  \BibitemOpen
  \bibfield  {author} {\bibinfo {author} {\bibfnamefont {N.~G.}\ \bibnamefont
  {Kara\c{c}ayl\i{}}} \emph {et~al.},\ }\href@noop {} {\  (\bibinfo {year}
  {2023})},\ \Eprint {http://arxiv.org/abs/2306.06316} {arXiv:2306.06316
  [astro-ph.CO]} \BibitemShut {NoStop}%
\bibitem [{\citenamefont {Jin}\ \emph {et~al.}(2022)\citenamefont {Jin} \emph
  {et~al.}}]{Jin:2022ddg}%
  \BibitemOpen
  \bibfield  {author} {\bibinfo {author} {\bibfnamefont {S.}~\bibnamefont
  {Jin}} \emph {et~al.},\ }\href@noop {} {\  (\bibinfo {year} {2022})},\
  \Eprint {http://arxiv.org/abs/2212.03981} {arXiv:2212.03981 [astro-ph.IM]}
  \BibitemShut {NoStop}%
\bibitem [{\citenamefont {Murgia}\ \emph {et~al.}(2018)\citenamefont {Murgia},
  \citenamefont {Ir\v{s}i\v{c}},\ and\ \citenamefont {Viel}}]{Murgia:2018now}%
  \BibitemOpen
  \bibfield  {author} {\bibinfo {author} {\bibfnamefont {R.}~\bibnamefont
  {Murgia}}, \bibinfo {author} {\bibfnamefont {V.}~\bibnamefont
  {Ir\v{s}i\v{c}}}, \ and\ \bibinfo {author} {\bibfnamefont {M.}~\bibnamefont
  {Viel}},\ }\href {\doibase 10.1103/PhysRevD.98.083540} {\bibfield  {journal}
  {\bibinfo  {journal} {Phys. Rev. D}\ }\textbf {\bibinfo {volume} {98}},\
  \bibinfo {pages} {083540} (\bibinfo {year} {2018})},\ \Eprint
  {http://arxiv.org/abs/1806.08371} {arXiv:1806.08371 [astro-ph.CO]}
  \BibitemShut {NoStop}%
\bibitem [{\citenamefont {Chabanier}\ \emph
  {et~al.}(2019{\natexlab{b}})\citenamefont {Chabanier}, \citenamefont
  {Millea},\ and\ \citenamefont {Palanque-Delabrouille}}]{Chabanier:2019eai}%
  \BibitemOpen
  \bibfield  {author} {\bibinfo {author} {\bibfnamefont {S.}~\bibnamefont
  {Chabanier}}, \bibinfo {author} {\bibfnamefont {M.}~\bibnamefont {Millea}}, \
  and\ \bibinfo {author} {\bibfnamefont {N.}~\bibnamefont
  {Palanque-Delabrouille}},\ }\href {\doibase 10.1093/mnras/stz2310} {\bibfield
   {journal} {\bibinfo  {journal} {Mon. Not. Roy. Astron. Soc.}\ }\textbf
  {\bibinfo {volume} {489}},\ \bibinfo {pages} {2247} (\bibinfo {year}
  {2019}{\natexlab{b}})},\ \Eprint {http://arxiv.org/abs/1905.08103}
  {arXiv:1905.08103 [astro-ph.CO]} \BibitemShut {NoStop}%
\bibitem [{\citenamefont {Cain}\ \emph {et~al.}(2023)\citenamefont {Cain},
  \citenamefont {D'Aloisio}, \citenamefont {Irsic}, \citenamefont {Gangolli},\
  and\ \citenamefont {Dhami}}]{Cain:2022ehj}%
  \BibitemOpen
  \bibfield  {author} {\bibinfo {author} {\bibfnamefont {C.}~\bibnamefont
  {Cain}}, \bibinfo {author} {\bibfnamefont {A.}~\bibnamefont {D'Aloisio}},
  \bibinfo {author} {\bibfnamefont {V.}~\bibnamefont {Irsic}}, \bibinfo
  {author} {\bibfnamefont {N.}~\bibnamefont {Gangolli}}, \ and\ \bibinfo
  {author} {\bibfnamefont {S.}~\bibnamefont {Dhami}},\ }\href {\doibase
  10.1088/1475-7516/2023/01/002} {\bibfield  {journal} {\bibinfo  {journal}
  {JCAP}\ }\textbf {\bibinfo {volume} {01}},\ \bibinfo {pages} {002} (\bibinfo
  {year} {2023})},\ \Eprint {http://arxiv.org/abs/2207.02876} {arXiv:2207.02876
  [astro-ph.CO]} \BibitemShut {NoStop}%
\bibitem [{\citenamefont {Ir\v{s}i\v{c}}\ \emph
  {et~al.}(2017{\natexlab{b}})\citenamefont {Ir\v{s}i\v{c}} \emph
  {et~al.}}]{Irsic:2017ixq}%
  \BibitemOpen
  \bibfield  {author} {\bibinfo {author} {\bibfnamefont {V.}~\bibnamefont
  {Ir\v{s}i\v{c}}} \emph {et~al.},\ }\href {\doibase
  10.1103/PhysRevD.96.023522} {\bibfield  {journal} {\bibinfo  {journal} {Phys.
  Rev. D}\ }\textbf {\bibinfo {volume} {96}},\ \bibinfo {pages} {023522}
  (\bibinfo {year} {2017}{\natexlab{b}})},\ \Eprint
  {http://arxiv.org/abs/1702.01764} {arXiv:1702.01764 [astro-ph.CO]}
  \BibitemShut {NoStop}%
\bibitem [{\citenamefont {{Neyman}}(1937)}]{Neyman}%
  \BibitemOpen
  \bibfield  {author} {\bibinfo {author} {\bibfnamefont {J.}~\bibnamefont
  {{Neyman}}},\ }\href {\doibase 10.1098/rsta.1937.0005} {\bibfield  {journal}
  {\bibinfo  {journal} {Philosophical Transactions of the Royal Society of
  London Series A}\ }\textbf {\bibinfo {volume} {236}},\ \bibinfo {pages} {333}
  (\bibinfo {year} {1937})}\BibitemShut {NoStop}%
\bibitem [{\citenamefont {Feldman}\ and\ \citenamefont
  {Cousins}(1998)}]{Feldman:1997qc}%
  \BibitemOpen
  \bibfield  {author} {\bibinfo {author} {\bibfnamefont {G.~J.}\ \bibnamefont
  {Feldman}}\ and\ \bibinfo {author} {\bibfnamefont {R.~D.}\ \bibnamefont
  {Cousins}},\ }\href {\doibase 10.1103/PhysRevD.57.3873} {\bibfield  {journal}
  {\bibinfo  {journal} {Phys. Rev. D}\ }\textbf {\bibinfo {volume} {57}},\
  \bibinfo {pages} {3873} (\bibinfo {year} {1998})},\ \Eprint
  {http://arxiv.org/abs/physics/9711021} {arXiv:physics/9711021} \BibitemShut
  {NoStop}%
\bibitem [{\citenamefont {Kirkpatrick}\ \emph {et~al.}(1983)\citenamefont
  {Kirkpatrick}, \citenamefont {Gelatt},\ and\ \citenamefont
  {Vecchi}}]{Kirkpatrick:1983zz}%
  \BibitemOpen
  \bibfield  {author} {\bibinfo {author} {\bibfnamefont {S.}~\bibnamefont
  {Kirkpatrick}}, \bibinfo {author} {\bibfnamefont {C.~D.}\ \bibnamefont
  {Gelatt}}, \ and\ \bibinfo {author} {\bibfnamefont {M.~P.}\ \bibnamefont
  {Vecchi}},\ }\href {\doibase 10.1126/science.220.4598.671} {\bibfield
  {journal} {\bibinfo  {journal} {Science}\ }\textbf {\bibinfo {volume}
  {220}},\ \bibinfo {pages} {671} (\bibinfo {year} {1983})}\BibitemShut
  {NoStop}%
\bibitem [{\citenamefont {Hannestad}(2000)}]{Hannestad:2000wx}%
  \BibitemOpen
  \bibfield  {author} {\bibinfo {author} {\bibfnamefont {S.}~\bibnamefont
  {Hannestad}},\ }\href {\doibase 10.1103/PhysRevD.61.023002} {\bibfield
  {journal} {\bibinfo  {journal} {Phys. Rev. D}\ }\textbf {\bibinfo {volume}
  {61}},\ \bibinfo {pages} {023002} (\bibinfo {year} {2000})},\ \Eprint
  {http://arxiv.org/abs/astro-ph/9911330} {arXiv:astro-ph/9911330} \BibitemShut
  {NoStop}%
\bibitem [{\citenamefont {Raveri}\ and\ \citenamefont
  {Hu}(2019)}]{Raveri:2018wln}%
  \BibitemOpen
  \bibfield  {author} {\bibinfo {author} {\bibfnamefont {M.}~\bibnamefont
  {Raveri}}\ and\ \bibinfo {author} {\bibfnamefont {W.}~\bibnamefont {Hu}},\
  }\href {\doibase 10.1103/PhysRevD.99.043506} {\bibfield  {journal} {\bibinfo
  {journal} {Phys. Rev. D}\ }\textbf {\bibinfo {volume} {99}},\ \bibinfo
  {pages} {043506} (\bibinfo {year} {2019})},\ \Eprint
  {http://arxiv.org/abs/1806.04649} {arXiv:1806.04649 [astro-ph.CO]}
  \BibitemShut {NoStop}%
\bibitem [{\citenamefont {Nicola}\ \emph {et~al.}(2019)\citenamefont {Nicola},
  \citenamefont {Amara},\ and\ \citenamefont {Refregier}}]{Nicola:2018rcd}%
  \BibitemOpen
  \bibfield  {author} {\bibinfo {author} {\bibfnamefont {A.}~\bibnamefont
  {Nicola}}, \bibinfo {author} {\bibfnamefont {A.}~\bibnamefont {Amara}}, \
  and\ \bibinfo {author} {\bibfnamefont {A.}~\bibnamefont {Refregier}},\ }\href
  {\doibase 10.1088/1475-7516/2019/01/011} {\bibfield  {journal} {\bibinfo
  {journal} {JCAP}\ }\textbf {\bibinfo {volume} {01}},\ \bibinfo {pages} {011}
  (\bibinfo {year} {2019})},\ \Eprint {http://arxiv.org/abs/1809.07333}
  {arXiv:1809.07333 [astro-ph.CO]} \BibitemShut {NoStop}%
\bibitem [{\citenamefont {Handley}\ and\ \citenamefont
  {Lemos}(2019)}]{Handley:2019wlz}%
  \BibitemOpen
  \bibfield  {author} {\bibinfo {author} {\bibfnamefont {W.}~\bibnamefont
  {Handley}}\ and\ \bibinfo {author} {\bibfnamefont {P.}~\bibnamefont
  {Lemos}},\ }\href {\doibase 10.1103/PhysRevD.100.043504} {\bibfield
  {journal} {\bibinfo  {journal} {Phys. Rev. D}\ }\textbf {\bibinfo {volume}
  {100}},\ \bibinfo {pages} {043504} (\bibinfo {year} {2019})},\ \Eprint
  {http://arxiv.org/abs/1902.04029} {arXiv:1902.04029 [astro-ph.CO]}
  \BibitemShut {NoStop}%
\bibitem [{\citenamefont {Hooper}\ and\ \citenamefont
  {Lucca}(2022)}]{Hooper:2021rjc}%
  \BibitemOpen
  \bibfield  {author} {\bibinfo {author} {\bibfnamefont {D.~C.}\ \bibnamefont
  {Hooper}}\ and\ \bibinfo {author} {\bibfnamefont {M.}~\bibnamefont {Lucca}},\
  }\href {\doibase 10.1103/PhysRevD.105.103504} {\bibfield  {journal} {\bibinfo
   {journal} {Phys. Rev. D}\ }\textbf {\bibinfo {volume} {105}},\ \bibinfo
  {pages} {103504} (\bibinfo {year} {2022})},\ \Eprint
  {http://arxiv.org/abs/2110.04024} {arXiv:2110.04024 [astro-ph.CO]}
  \BibitemShut {NoStop}%
\bibitem [{\citenamefont {Hooper}\ \emph {et~al.}(2022)\citenamefont {Hooper},
  \citenamefont {Sch\"oneberg}, \citenamefont {Murgia}, \citenamefont
  {Archidiacono}, \citenamefont {Lesgourgues},\ and\ \citenamefont
  {Viel}}]{Hooper:2022byl}%
  \BibitemOpen
  \bibfield  {author} {\bibinfo {author} {\bibfnamefont {D.~C.}\ \bibnamefont
  {Hooper}}, \bibinfo {author} {\bibfnamefont {N.}~\bibnamefont
  {Sch\"oneberg}}, \bibinfo {author} {\bibfnamefont {R.}~\bibnamefont
  {Murgia}}, \bibinfo {author} {\bibfnamefont {M.}~\bibnamefont
  {Archidiacono}}, \bibinfo {author} {\bibfnamefont {J.}~\bibnamefont
  {Lesgourgues}}, \ and\ \bibinfo {author} {\bibfnamefont {M.}~\bibnamefont
  {Viel}},\ }\href {\doibase 10.1088/1475-7516/2022/10/032} {\bibfield
  {journal} {\bibinfo  {journal} {JCAP}\ }\textbf {\bibinfo {volume} {10}},\
  \bibinfo {pages} {032} (\bibinfo {year} {2022})},\ \Eprint
  {http://arxiv.org/abs/2206.08188} {arXiv:2206.08188 [astro-ph.CO]}
  \BibitemShut {NoStop}%
\end{thebibliography}%

\clearpage

\pagebreak

%TC:ignore
\section{Supplemental material}

\subsection{Details of Ly$\alpha$ likelihoods}

In this work, we use compressed Ly$\alpha$ forest likelihoods that model the 1D Ly$\alpha$ forest flux power spectrum by the dimensionless amplitude $\Delta_L^2\equiv k_p^3P_{\rm lin}(k_p,z_p)/(2\pi^2)$ and the logarithmic derivative $n_L\equiv \left( d\ln P_{\rm lin}(k, z)/d\ln k \right) |_{(k_p,z_p)}$ of the linear power spectrum $P_{\rm lin}$, both evaluated at a pivot redshift $z_p=3$ and wavenumber $k_p=0.009~{\rm s/km}$ \cite{SDSS:2004aee}. Here we present a series of results that validate both our implementation of the likelihood and its application to EDE cosmologies.

\begin{figure}[!t]
\includegraphics[width=0.99\linewidth]{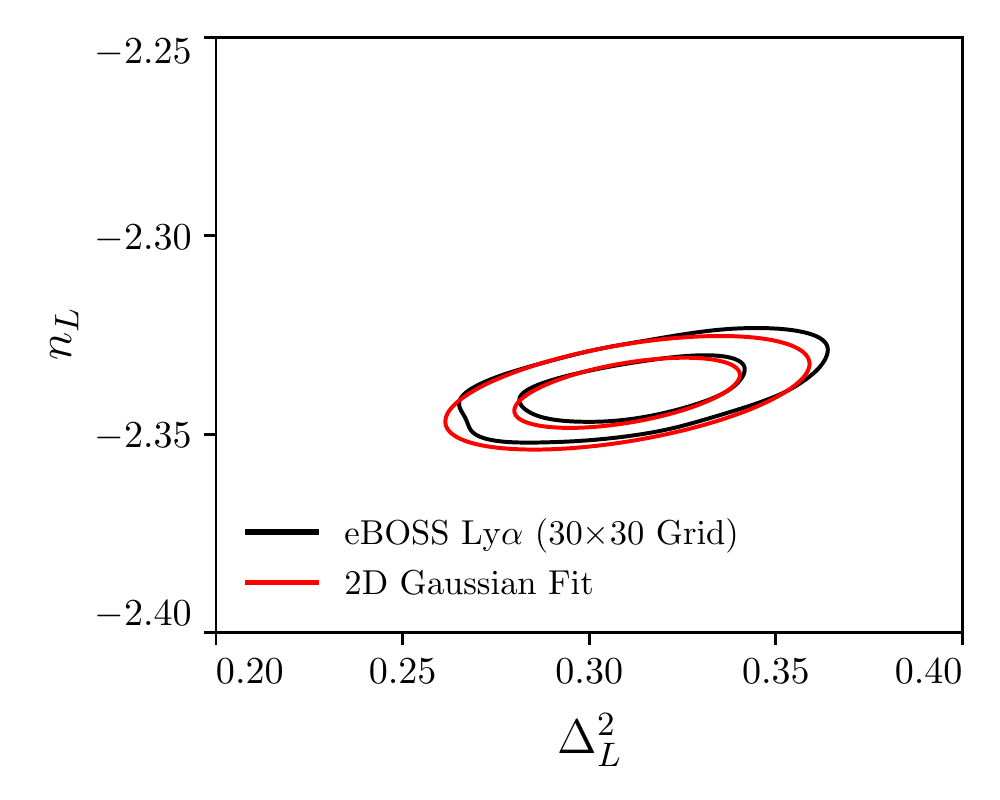}
\caption{Comparison of 2D confidence contour in the $\Delta_L^2$ -- $n_L$ plane derived from interpolating between a $30\times 30$ grid of samples from Fig.~20 of \cite{Chabanier:2018rga} (black) and from fitting the samples with a 2D Gaussian. Contours indicate the 68\% and 95\% CL regions.}
 \label{fig:gauss_fit}
\end{figure}

\begin{table}[!t]
    \centering
    {\scriptsize
    \begin{tabular}{ |c|c|c|c|c|c|c|c|c| }
     \hline
        & $\log(10^{10}A_s)$  &  $n_s$ &   $H_0$ & $\Omega_ch^2$  & $\Omega_bh^2$  & $\Delta _L^2$ & $n_L$ \\ 
      \hline
      $\Lambda$CDM & $~3.010~$ & $~0.9782~$ &  $~69.82~$ & $~0.1267~$ & $~0.0233~$  & 0.390& -2.285 \\ 
     \hline 
     EDE &   $~3.055~$  & $~0.9791~$ & $~70.33~$ & $~0.1267~$ & $~0.0226~$  & 0.387  &  -2.283
      \\ 
     \hline
     
    \end{tabular}
    }
    \caption{$\Lambda$CDM parameters for the reference cosmology used in Fig.~\ref{fig:Pk_lcdm_EDE}. $H_0$ values are in [km/s/Mpc] and the cosmology assumes massless neutrinos. For reference, we include the $\Lambda$CDM parameters for the MAP EDE cosmology from the baseline analysis and the effective Ly$\alpha$ parameters used in the compressed likelihoods. The EDE cosmology has $f_{\rm EDE}=0.072$, $\log_{10}(z_c)=3.56$, and $\theta_i=2.73.$}
   \label{tab:mimic_LCDM_params}
\end{table}

In order to reconstruct a continuous eBOSS Ly$\alpha$ likelihood from a sparsely sampled grid of $\Delta_L^2$ and $n_L$ values from~\cite{Chabanier:2018rga}, we model the contour with a 2D Gaussian. Fig.~\ref{fig:gauss_fit} compares the eBOSS Ly$\alpha$ confidence contour reconstructed from a $30\times30$ grid with the 2D Gaussian fit used in this work. The 2D Gaussian provides an excellent description of the 68\% and 95\% CL contours, and is thus a reliable model for the compressed eBOSS Ly$\alpha$ likelihood.

\begin{figure}[!t]
\includegraphics[width=0.99\linewidth]{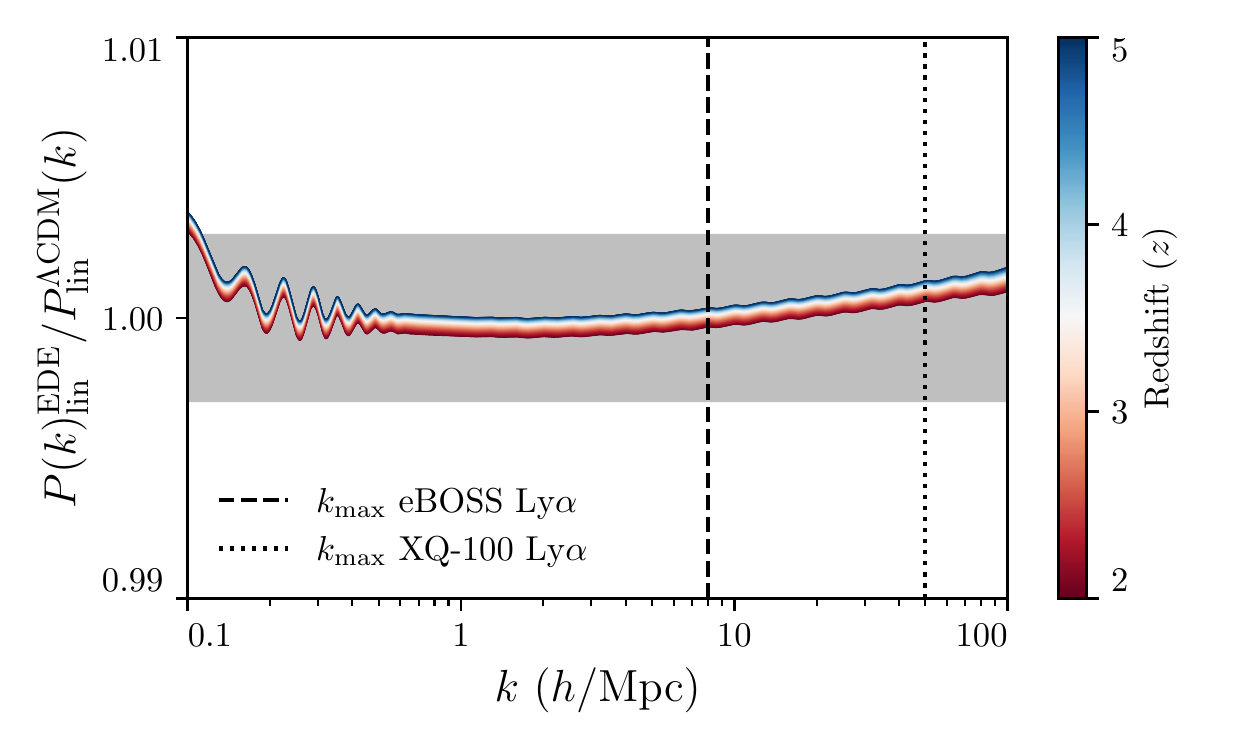}
\caption{Ratio of the linear matter power spectrum at redshifts $2\leq z\leq 5$ for a $\Lambda$CDM cosmology (with tuned parameters given in Table~\ref{tab:mimic_LCDM_params}) to that of the best-fit EDE cosmology from the baseline dataset. The vertical lines indicate the approximate maximum 3D wavenumber probed by each of the Ly$\alpha$ forest datasets~\cite{Murgia:2018now, Chabanier:2019eai, Cain:2022ehj}. The $\Lambda$CDM power spectrum agrees with the EDE power spectrum  to within 0.3\% (shaded grey) over the range of wavenumbers and redshifts probed by the Ly$\alpha$ forest measurements.}
 \label{fig:Pk_lcdm_EDE}
\end{figure}

\begin{figure}[!t]
\includegraphics[width=0.99\linewidth]{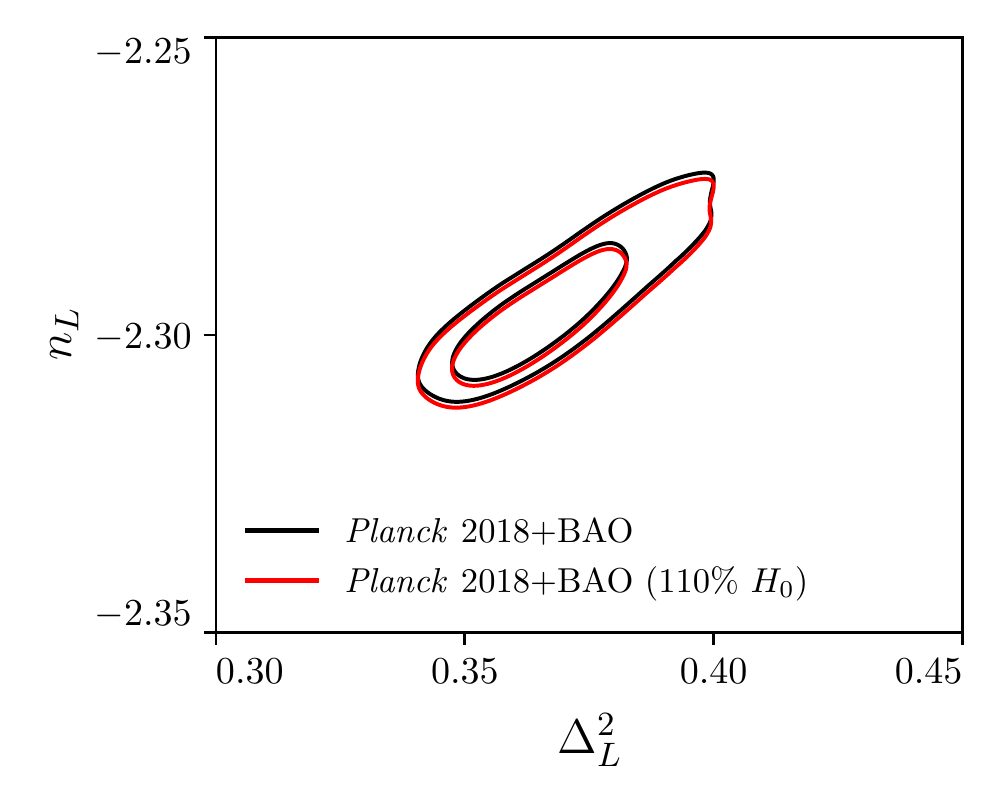}
\caption{Comparison of baseline dataset constraints on EDE ($n=3$) in the Ly$\alpha$ parameter space, with constraints after increasing $H_0$ by 10\% while leaving $f_{\rm EDE}$, $\log_{10}(z_c)$, $\theta_i$, $\Omega_bh^2$, $\Omega_ch^2$, $A_s$, $n_s$, and $\tau$ unchanged for each sample and re-computing $n_L$ and $\Delta_L^2$. Contours indicate the 68\% and 95\% CL regions.  It is evident that the compressed parameters $n_L$ and $\Delta_L^2$ are insensitive to the choice of $H_0$ prior used in the Ly$\alpha$ analysis.  }
 \label{fig:H0_validation}
\end{figure}

\begin{figure*}[!t]
\includegraphics[width=\linewidth]{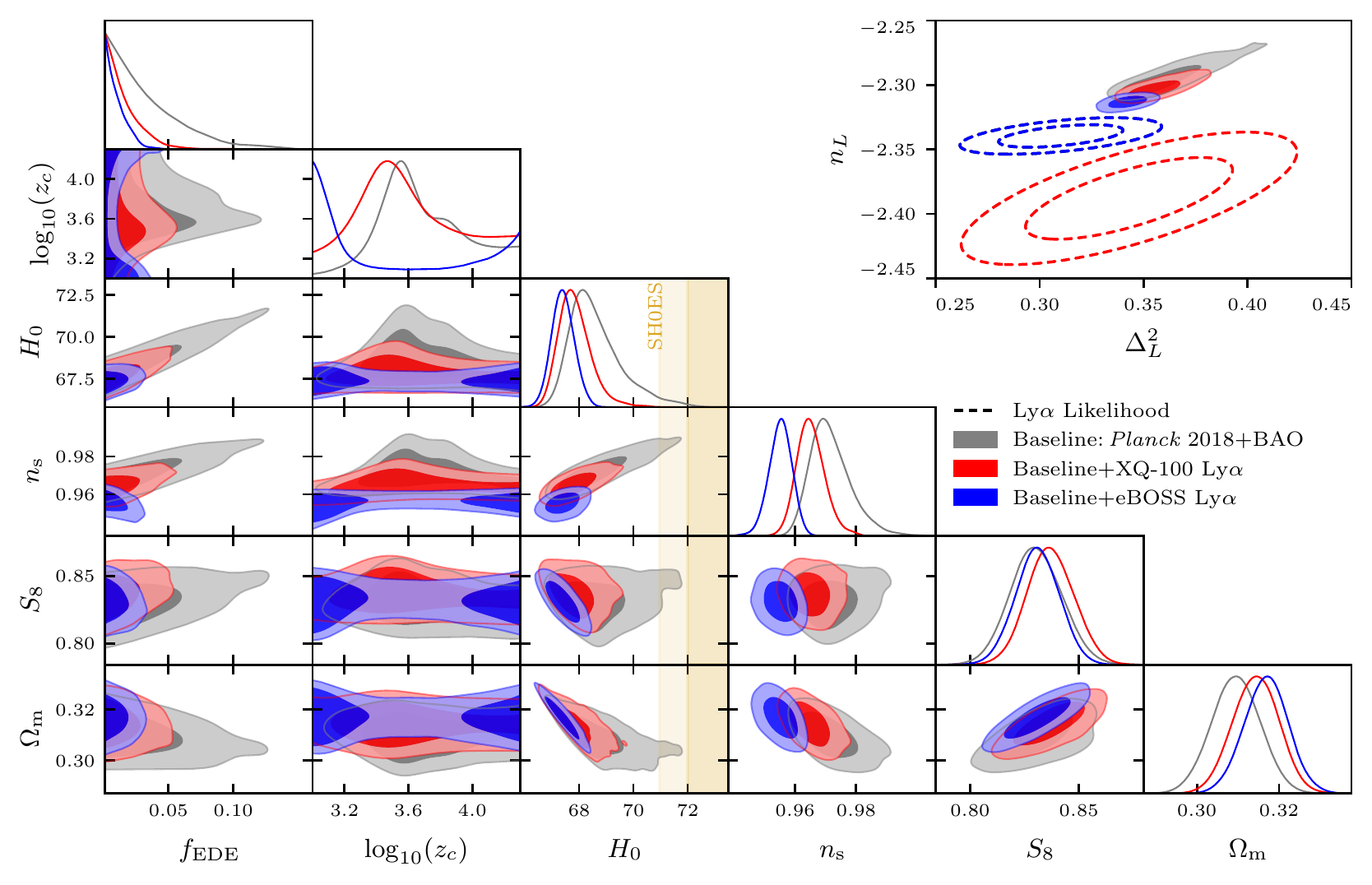}
\caption{Marginalized posteriors for a subset of EDE ($n=3$) and $\Lambda$CDM parameters before and after including Ly$\alpha$ data, assuming a single species of massive neutrinos with $\sum m_\nu = 0.06$ eV. Our results are effectively the same as those of the massless neutrino analysis in Fig.~\textcolor{Maroon}{2} of the main text.  }
 \label{fig:ly_massive_neutrinos}
\end{figure*}

\begin{figure*}
\includegraphics[width=\linewidth]{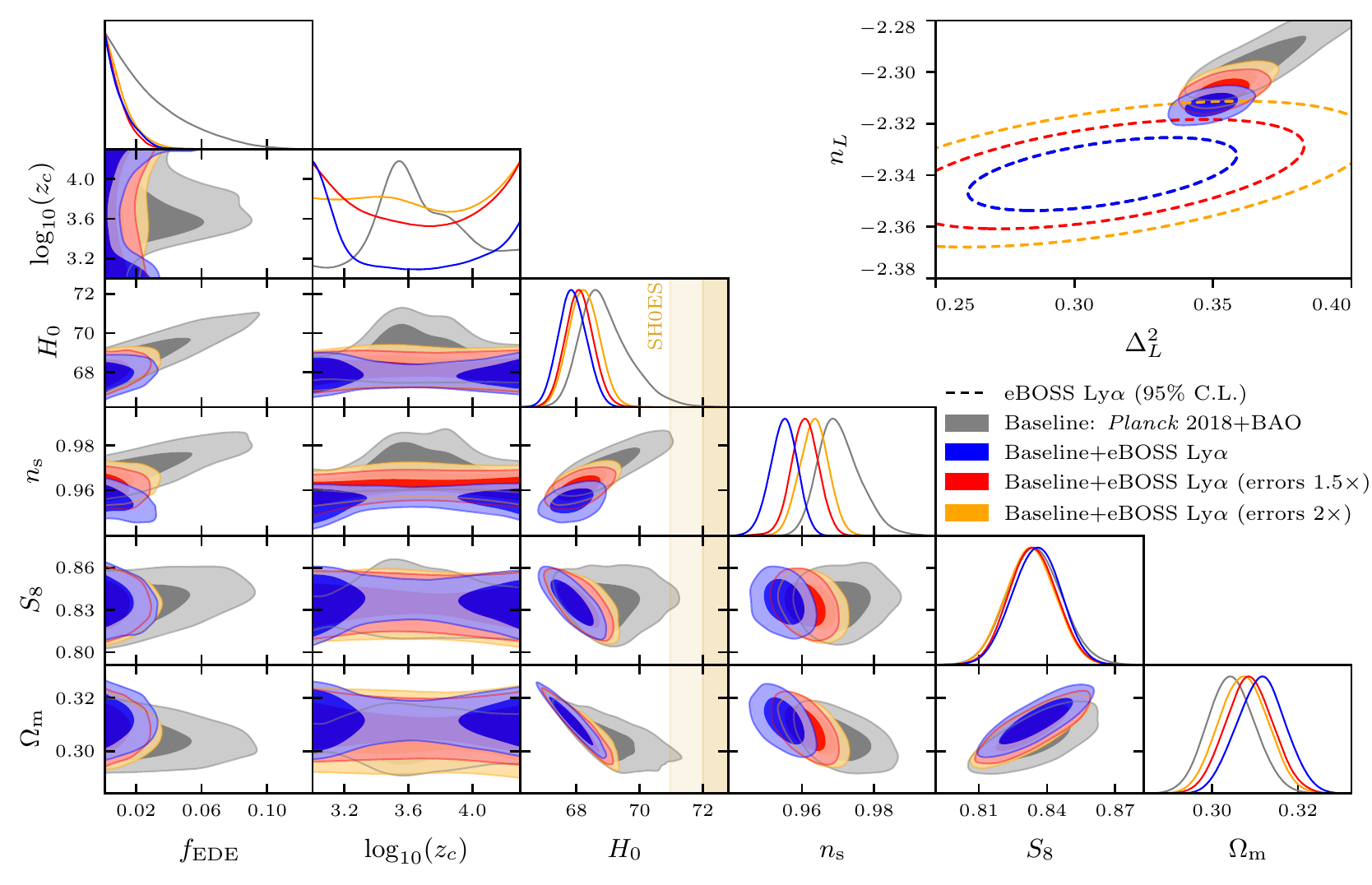}
\caption{Marginalized posteriors for a subset of EDE ($n=3$) and $\Lambda$CDM parameters for the baseline and baseline + eBOSS Ly$\alpha$ datasets with increased values of $\sigma_{\Delta_L^2}$ and $\sigma_{n_L}$ (as labeled in the legend) for the Ly$\alpha$ likelihood. The top right panel shows constraints in the $\Delta_L^2$ -- $n_L$ plane, as well as the 95\% C.L. for the modified eBOSS Ly$\alpha$ likelihood. The main effect of increasing $\sigma_{\Delta_L^2}$ and $\sigma_{n_L}$ is an increase in $n_s$; however, even with an increase of $\sigma_{\Delta_L^2}$ and $\sigma_{n_L}$ by a factor of two, the eBOSS Ly$\alpha$ likelihood rules out Hubble-tension-resolving EDE cosmologies.  } 
 \label{fig:ly_alpha_error_increase}
\end{figure*}

The XQ-100 Ly$\alpha$ likelihood is constructed by fitting a 2D Gaussian to samples of $n_L$ and $\Delta_L^2$ derived from the analysis in Appendix A of~\cite{Esposito:2022plo}. In particular, the analysis in~\cite{Esposito:2022plo} varies astrophysical IGM parameters, as well as the amplitude $\sigma_8$ and the slope of the linear matter power spectrum $n_{\rm eff}\equiv d\ln P(k)/d\ln(k)$ evaluated at $k=0.005~$s/km, with the remaining cosmological parameters fixed to \emph{Planck}-like values. We use the chain from this work to derive samples in the $\Delta_L^2$ -- $n_L$ plane, which we fit with a 2D Gaussian. Although the XQ-100 Ly$\alpha$ measurements include smaller scales than the eBOSS Ly$\alpha$ measurements, the compressed likelihood, which is evaluated at the same pivot scale and redshift as was used in the eBOSS analysis, still provides an accurate description of the cosmological information in the 1D Ly$\alpha$ forest flux power spectrum in cosmologies without sharp features in the small-scale matter power spectrum. This is because the XQ-100 Ly$\alpha$ likelihood marginalizes over a wide range of thermal histories at small scales~\cite{Irsic:2017ixq}, and the cosmological information from the XQ-100 Ly$\alpha$ forest measurements is well-described by the late-time amplitude and slope of the linear matter power spectrum evaluated at a pivot scale, as shown in~\cite{Esposito:2022plo}. 

Since both likelihoods marginalize over baryonic physics by running grids of hydrodynamical simulations, the only assumption we make is that these grids, which are run in $\Lambda$CDM parameter space, also cover relevant $P_{\rm lin}(k)$ for our EDE model fits over the range of wavenumbers and redshifts to which the Ly$\alpha$ forest 1D flux power spectrum is sensitive.  To validate this assumption, we explicitly check that $P_{\rm lin}(k)$ in cosmologies with non-negligible EDE can be accurately mimicked by $P_{\rm lin}(k)$ in a $\Lambda$CDM cosmology with different cosmological parameters (\textit{e.g.}, $n_s$, $h$, etc.).  Fig.~\ref{fig:Pk_lcdm_EDE} shows the ratio of the linear matter power spectrum across redshifts $2 \leq z \leq 5$ for a reference (mimic) $\Lambda$CDM cosmology (parameters listed in Table~\ref{tab:mimic_LCDM_params}) to that of the best-fit EDE cosmology from the baseline dataset (MAP parameters from the second row of Table~\ref{tab:posterior_full_supp_material} 
of the Supplemental Material and listed in Table~\ref{tab:mimic_LCDM_params} for convenience). The two agree within 0.3\% over the range of scales and redshifts probed by the 1D Ly$\alpha$ forest flux power spectrum~\cite{Murgia:2018now, Chabanier:2019eai, Cain:2022ehj}. Thus, $P_{\rm lin}(k)$ for the EDE cosmologies considered here can be very accurately captured by a $\Lambda$CDM $P_{\rm lin}(k)$ with tuned parameters that fall within the region of parameter space used in the eBOSS and XQ-100 analyses. This validates our use of the eBOSS and XQ-100 Ly$\alpha$ likelihoods, as currently constructed.  A complete validation of the Ly$\alpha$ likelihoods in the EDE context requires running a grid of EDE-based hydrodynamical simulations, which is beyond the scope of this work. 

An additional subtlety is that the eBOSS Ly$\alpha$ likelihood uses a tight prior on $H_0$, as this allows the $\Omega_m$ -- $H_0$ degeneracy in $P(k)$ to be broken and, hence, constraints to be derived on the matter density. However, such priors on $H_0$ actually have no impact on the compressed Ly$\alpha$ parameters $n_L$ and $\Delta_L^2$, and hence do not affect our results.  As a test of the dependence of our results on the $H_0$ priors used in the Ly$\alpha$ likelihoods, we recompute the baseline dataset EDE constraints on $n_L$ and $\Delta_L^2$ after increasing $H_0$ by 10\% for each sample in the chain while leaving $f_{\rm EDE}$, $\log_{10}(z_c)$, $\theta_i$, $\Omega_bh^2$, $\Omega_ch^2$, $A_s$, $n_s$, and $\tau$ fixed. The results are shown in Fig.~\ref{fig:H0_validation}. Constraints on $n_L$ and $\Delta_L^2$ are clearly insensitive to the value of $H_0$, and thus to the $H_0$ prior. See Ref.~\cite{Pedersen:2022anu} for a more detailed discussion of the impact of priors on Ly$\alpha$ constraints in the $\Delta_L^2$ -- $n_L$ plane.

Finally, since the eBOSS Ly$\alpha$ likelihood assumes massless neutrinos~\cite{Chabanier:2018rga} and the XQ-100 likelihood assumes a single species of massive neutrinos with $\sum m_\nu = 0.06$ eV~\cite{Esposito:2022plo}, we re-analyze the baseline and baseline + Ly$\alpha$ datasets for an EDE cosmology assuming the sum of the neutrino masses $\sum m_\nu = 0.06$ eV, with one massive eigenstate and two massless eigenstates. Fig.~\ref{fig:ly_massive_neutrinos} shows the marginalized posteriors for this analysis. Although including a single species of neutrinos with non-zero mass leads to slight shifts in the inferred parameter values, our conclusions are unchanged.

\subsection{Robustness of Ly$\alpha$-based EDE analyses}

\begin{figure*}[!t]
\includegraphics[width=\linewidth]{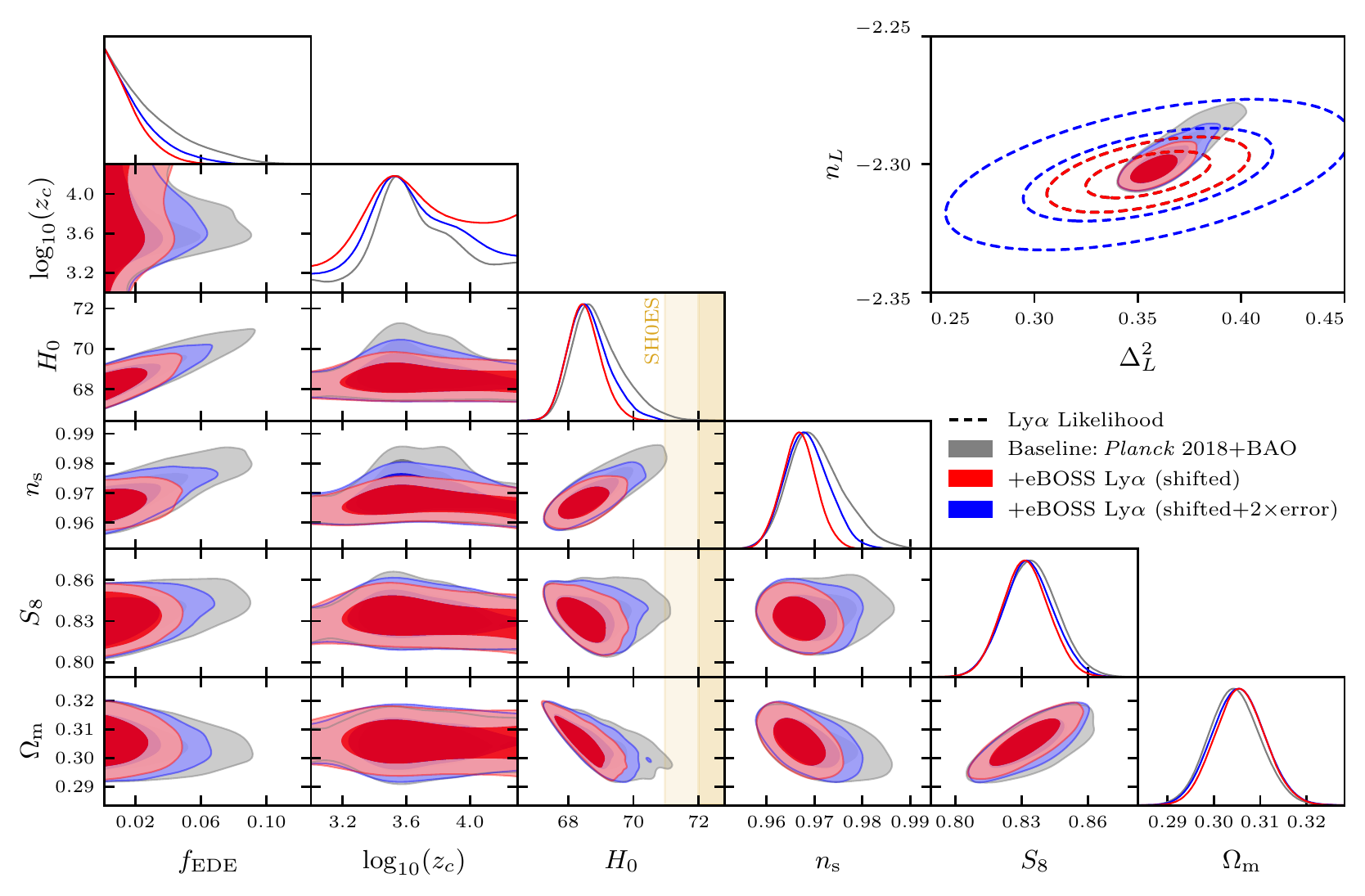}
\caption{Marginalized posteriors for a subset of EDE ($n=3$) and $\Lambda$CDM parameters for the baseline dataset before and after including a ``shifted" eBOSS Ly$\alpha$ likelihood centered at the posterior mean values $\bar{\Delta}_L^2$ and $\bar{n}_L$ from the baseline $\Lambda$CDM analysis. Including the shifted likelihood (red) significantly constrains the EDE parameter space, demonstrating that even if the eBOSS data were not in tension within $\Lambda$CDM, they would significantly constrain EDE models. The shifted likelihood strongly constrains the EDE parameter space even if we double $\sigma_{\Delta_L^2}$ and $\sigma_{n_L}$ (blue).}
 \label{fig:ly_alpha_shift_eBOSS}
\end{figure*}

\begin{figure*}[!t]
\includegraphics[width=\linewidth]{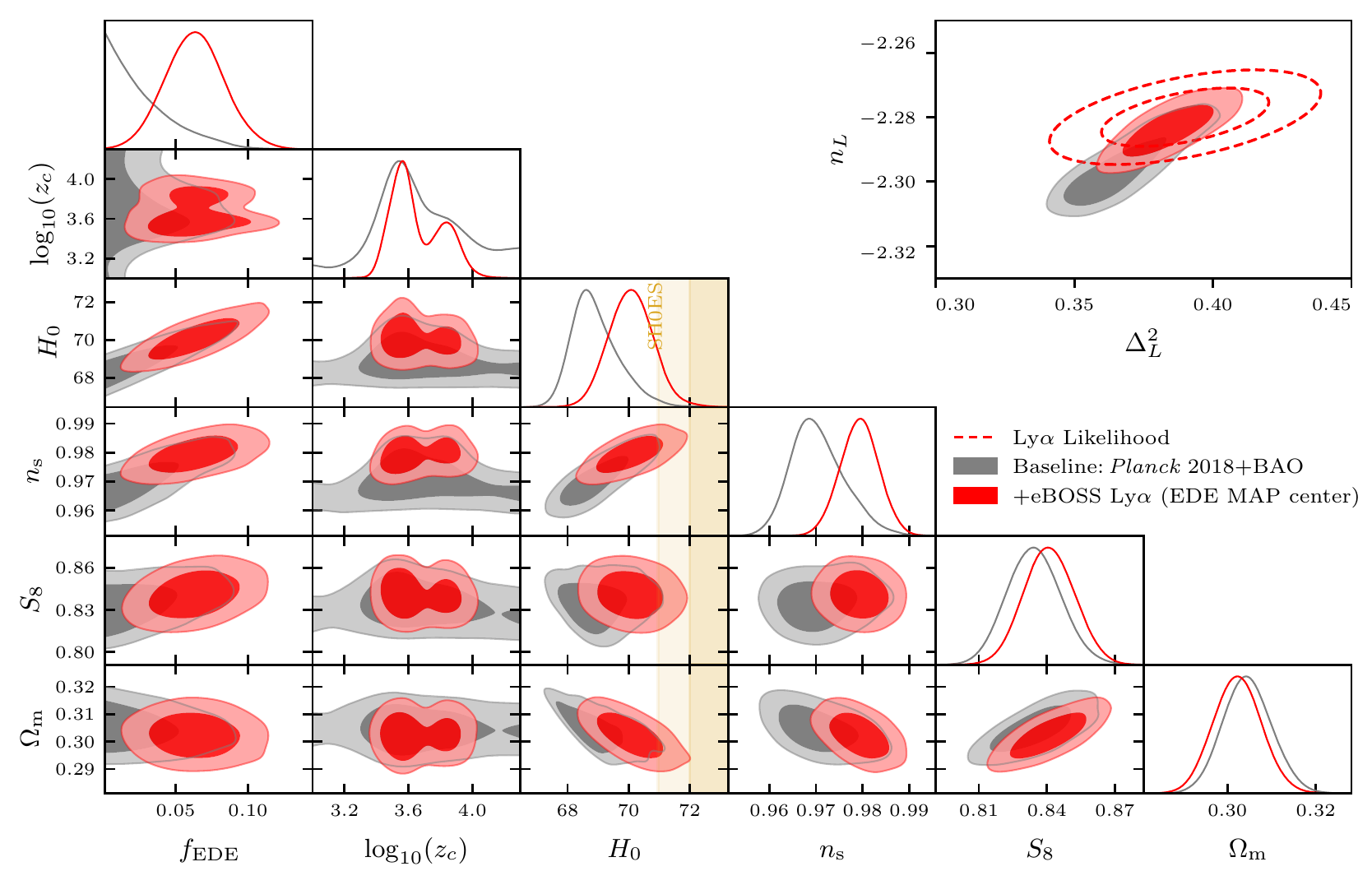}
\caption{Marginalized posteriors for a subset of EDE ($n=3$) and $\Lambda$CDM parameters for the baseline dataset before and after including a ``shifted" eBOSS Ly$\alpha$ likelihood centered near the MAP values of $\Delta_L^2$ and $n_L$ from the baseline EDE analysis. Including the shifted likelihood (red) leads to a preference for non-zero values of $f_{\rm EDE}$, demonstrating that if future Ly$\alpha$ forest analyses find large values of $\Delta_L^2$ and $n_L$, then they could provide compelling evidence for EDE.}
 \label{fig:ly_alpha_EDE_fake}
\end{figure*}

Given the tension between the \emph{Planck} + BAO and Ly$\alpha$ data, we perform a series of tests to assess the robustness of our EDE analyses to potential systematics in the compressed Ly$\alpha$ likelihoods and to test the extent to which our tight constraints on the EDE parameter space are driven by the existing tension within $\Lambda$CDM.

We first assess how sensitive our conclusions are to the precision of the Ly$\alpha$ forest likelihoods. Since the baseline + eBOSS Ly$\alpha$ dataset is the most constraining dataset combination considered in this work, we re-analyze this combination with the eBOSS uncertainties $\sigma_{\Delta_L^2}$ and $\sigma_{n_L}$ both inflated, first by 50\% and then by 100\%. Fig.~\ref{fig:ly_alpha_error_increase} shows the marginalized posteriors for a subset of the EDE and $\Lambda$CDM parameters for these datasets. The top right panel compares constraints on $n_L$ and $\Delta_L^2$ with the 95\% limits for the modified eBOSS Ly$\alpha$ likelihoods. Whereas doubling $\sigma_{n_L}$ and $\sigma_{\Delta_L^2}$ significantly reduces the tension between the baseline constraints and the eBOSS Ly$\alpha$ likelihood in the $\Delta_L^2$ -- $n_L$ plane, it has little impact on the $f_{\rm EDE}$ and $H_0$ posteriors. This can be understood by noticing that the main impact of increasing  $\sigma_{n_L}$ and $\sigma_{\Delta_L^2}$ is an increase in $n_s$; however, $n_s$ needs to be sufficiently large in order to have an EDE cosmology that is both compatible with CMB data and has a large value of $f_{\rm EDE}$. 

Second, we assess the sensitivity of our conclusions to shifts in the central value of the eBOSS Ly$\alpha$ forest likelihood. Fig.~\ref{fig:ly_alpha_shift_eBOSS} shows the marginalized posteriors for a subset of the EDE and $\Lambda$CDM parameters for the baseline dataset, as well as for the baseline dataset with a ``shifted" eBOSS Ly$\alpha$ likelihood centered at the posterior mean of the baseline $\Lambda$CDM analysis $(\bar\Delta_{L^2}, \bar{n}_{L})=(0.355, -2.304)$. Even this shifted likelihood (red) significantly reduces the allowed values of $H_0$ and $f_{\rm EDE}$ compared to the baseline analysis and, as discussed in the main text, provides even tighter constraints on $f_{\rm EDE}$ and $H_0$ than the XQ-100 Ly$\alpha$ likelihood. These tight constraints are driven by the precision of the eBOSS likelihood and the misalignment between the $\Delta_L^2$ -- $n_L$ degeneracy axis of the eBOSS likelihood and that of the baseline analysis. The constraints weaken if we both shift the mean and double the eBOSS uncertainties (blue); nevertheless, even this highly modified likelihood places significant constraints on the EDE parameter space.\footnote{It is worth noting that the analysis including the shifted and error-inflated likelihood is affected by prior-volume effects similar to those in the baseline analysis.} This exercise demonstrates that even if the eBOSS Ly$\alpha$ likelihood were not in tension with the baseline $\Lambda$CDM analysis, it would still place stringent constraints on EDE cosmologies capable of resolving the Hubble tension. 

Finally, we investigate whether a Ly$\alpha$ forest measurement with the same precision as the eBOSS Ly$\alpha$ likelihood, but larger values of $\Delta_L^2$ and $n_L$, could lead to a significant detection of $f_{\rm EDE}.$ To this end, we shift the central value of the eBOSS Ly$\alpha$ likelihood to  $(\bar\Delta_{L^2}, \bar{n}_{L})=(0.39, -2.28)$, near the MAP of the baseline EDE analysis (Table~\ref{tab:mimic_LCDM_params}). Fig.~\ref{fig:ly_alpha_EDE_fake} compares the marginalized posteriors for the baseline dataset with those derived from an analysis including the baseline dataset and this shifted likelihood. Including the shifted likelihood leads to a ``detection" of $f_{\rm EDE}=0.063^{+0.021}_{-0.021}$ at 68\% CL and an increase in the $H_0$ posterior. Consequently, if future analyses of the Ly$\alpha$ forest find values of $n_L$ and $\Delta_L^2$ (much) larger than those favored by current constraints, then these datasets could provide compelling evidence for EDE when analyzed in tandem with CMB and BAO data. Nevertheless, seeing as current Ly$\alpha$ forest analyses report low values of $n_L$ and $\Delta_L^2$ relative to \emph{Planck} $\Lambda$CDM, this scenario would indicate the presence of significant systematics in the Ly$\alpha$ datasets considered here.

\subsection{Profile likelihood analysis}

In this section, we use a profile likelihood to assess the impact of prior-volume effects on the MCMC constraints presented in the main text. MCMC analyses of EDE cosmologies are particularly susceptible to prior-volume effects because, at low values of $f_{\rm EDE}$, changes in the EDE parameters $\log_{10}(z_c)$ and $\theta_i$ have minimal discernible impact on the likelihood~\cite{Smith:2020rxx, Herold:2021ksg, Gomez-Valent:2022hkb, Herold:2022iib}. Consequently, the marginalized posteriors for $\log_{10}(z_c)$ and $\theta_i$ become prior-dominated at low values of $f_{\rm EDE}$ (see the baseline constraints on $f_{\rm EDE}$--$\log_{10}(z_c)$ in Fig.~\ref{fig:ly_alpha_EDE_fake} for an example), which can lead to a preference for low values of $f_{\rm EDE}$ in the marginalized posterior. As shown in~\cite{Herold:2021ksg, Herold:2022iib}, the profile likelihood is a powerful tool to investigate the impact of prior-volume effects on MCMC constraints for EDE cosmologies and to derive confidence limits on $f_{\rm EDE}$ and $H_0$ that are independent of priors.\footnote{A complementary approach to mitigate prior-volume effects in EDE constraints is the one-parameter EDE model presented in~\cite{Smith:2020rxx}, where $f_{\rm EDE}$ is the only free EDE parameter. As shown in~\cite{Herold:2021ksg}, this model is highly sensitive to the choice of $\log_{10}(z_c)$ and $\theta_i$. Hence, we utilize a profile likelihood instead.} However, it is worth noting that the profile likelihood approach imposes no penalty for an excessive number of free parameters in the model, \emph{i.e.}, it does not disfavor fine-tuning, whereas the prior-volume effect described above effectively disfavors excessive fine-tuning in MCMC analyses.

\begin{figure}[!tbp]

\includegraphics[width=\linewidth]{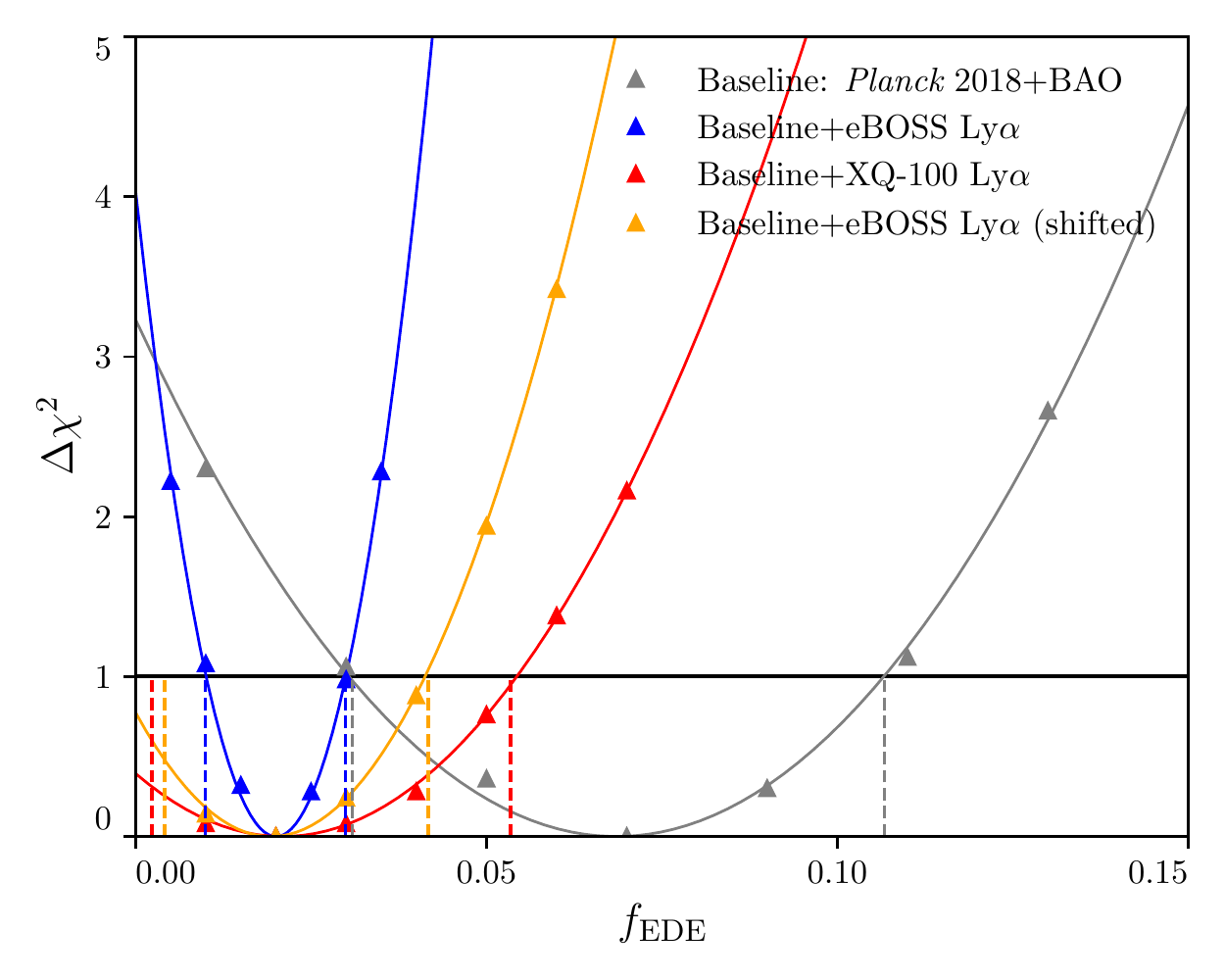}
\caption{Profile likelihoods for $f_{\rm EDE}$ for the dataset combinations considered in the main text (see Fig.~2 of the main text). Solid lines represent parabolic fits to the $\Delta\chi^2(f_{\rm EDE})$ values, and the black horizontal line corresponds to $\Delta\chi^2(f_{\rm EDE})=1$. The intersection between the black line and a parabolic fit indicates the 68.27\% confidence interval using the Neyman prescription~\cite{Neyman}. The dashed vertical lines denote the 68.27\% Feldman-Cousins confidence interval~\cite{Feldman:1997qc}. Including Ly$\alpha$ forest measurements significantly reduces the upper bound on $f_{\rm EDE}.$ The fact that the profile likelihoods including Ly$\alpha$ data are centered around $f_{\rm EDE}=0.02$ is a coincidence.}
 \label{fig:profile_likelihood}
\end{figure}

Our profile likelihood implementation is based on the analysis in~\cite{Herold:2021ksg}. We construct a profile likelihood for $f_{\rm EDE}$ by maximizing the log-likelihood, $\log\mathcal{L}(f_{\rm EDE})$ (\emph{i.e.}, minimizing $\chi^2(f_{\rm EDE})=-2\log\mathcal{L}(f_{\rm EDE})$) at fixed values of $f_{\rm EDE}$.\footnote{Strictly speaking, we maximize the log-posterior instead of the log-likelihood. This distinction has negligible impact on our results since we assume wide, uniform priors on all cosmological parameters; nevertheless, we maximize the posterior so that we self-consistently include the information from the priors on the \emph{Planck} nuisance parameters used in the MCMC analyses.} Given the numerical difficulty of determining the maximum-likelihood point for EDE models using traditional descent-based algorithms, we determine the maximum-likelihood point and all MAP values quoted in the text using a simulated annealing~\cite{Kirkpatrick:1983zz, Hannestad:2000wx} approach based on the method outlined in~\cite{Schoneberg:2021qvd}. In particular, for a given value of $f_{\rm EDE}$, we run a long MCMC chain until $|R-1|\leq 0.2$, which allows us to estimate the covariance. We then run this chain at a lower temperature and a smaller proposal scale until approximately 2000 steps are accepted. Afterward, we further decrease the temperature and proposal scale,\footnote{In practice, we found that logarithmically decreasing the temperature and proposal scale works well.} repeating this process until we observe that the improvement in the maximum $\log\mathcal{L}(f_{\rm EDE})$ between successive chains with different temperature and proposal scales is less than $0.1$. The maximum-likelihood point then corresponds to the maximum-likelihood sample from this chain. 
This method should reliably produce $\Delta\chi^2(f_{\rm EDE})$ values that are accurate to within $\approx0.2,$ which is sufficient to construct a profile likelihood. For each dataset combination, we determine the maximum-likelihood point for seven different values of $f_{\rm EDE}$. This set of values allows us to reliably fit the $\Delta\chi^2(f_{\rm EDE})$ points with a two-parameter parabolic model, as discussed below.

\begin{figure*}
\includegraphics[width=\linewidth]{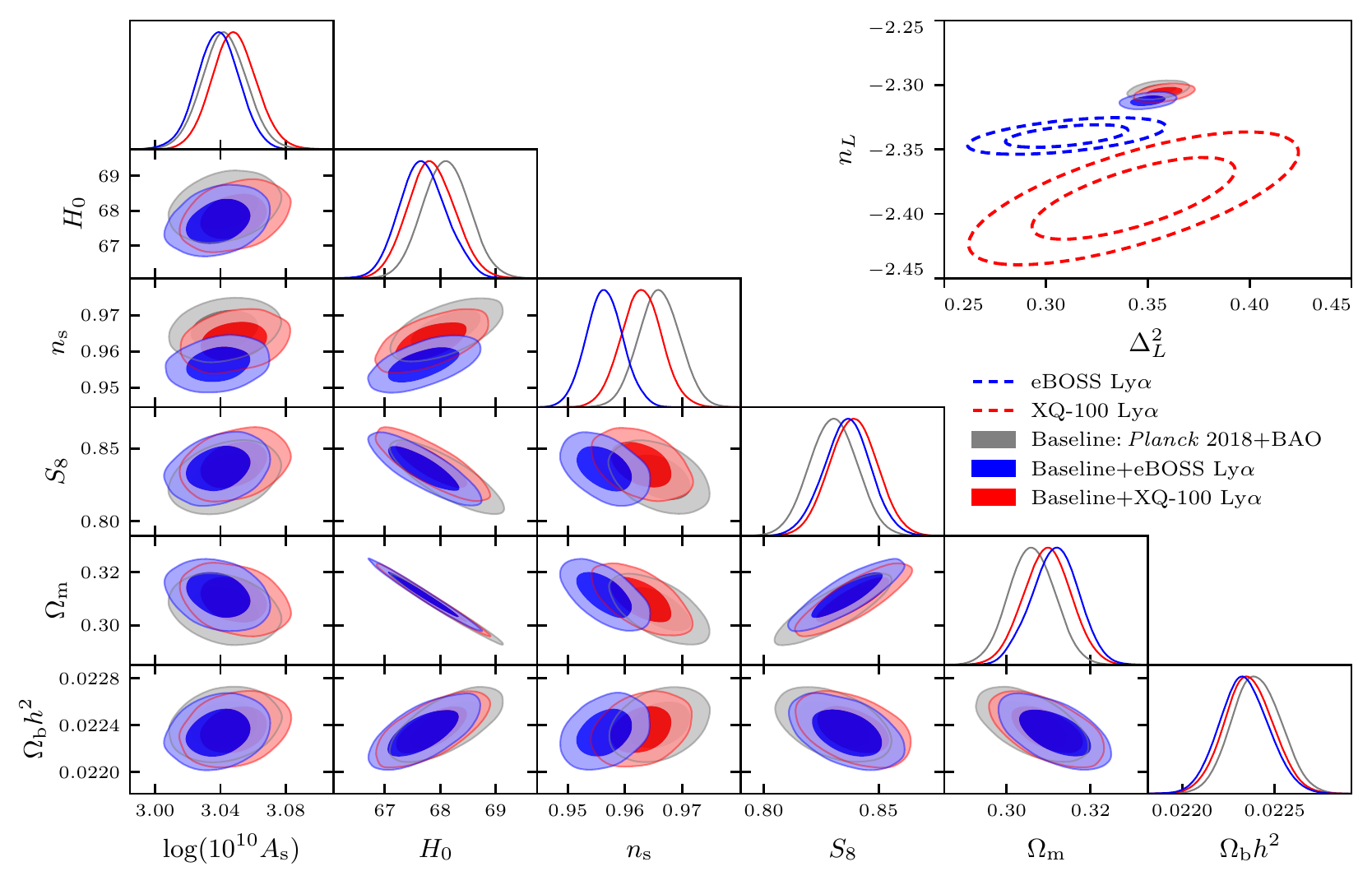}
\caption{Marginalized parameter posteriors for the $\Lambda$CDM analyses of the baseline and baseline + Ly$\alpha$ datasets. The top right panel shows constraints in the $\Delta_{L^2}$ -- $n_L$ plane for each of these datasets, as well as the Ly$\alpha$ likelihoods (dashed lines). The Ly$\alpha$ datasets are in moderate tension with the $\Lambda$CDM analysis of the baseline dataset.}
 \label{fig:ly_alpha_LCDM_constraints}
\end{figure*}

\begin{table}[!t]
    \centering
    {\scriptsize
    \begin{tabular}{ |c|c|c| }
     \hline
        Dataset  &  $f_{\rm EDE}$ (MCMC)   & $f_{\rm EDE}$ (Profile Likelihood)  \\ 
      \hline
      Baseline & $<0.073$ & $0.069_{-0.062}^{+0.074}$ \\
      \hline
      +eBOSS Ly$\alpha$ & $<0.028$ &  $0.020^{+0.019}_{-0.017}$ \\ 
      \hline
      +XQ-100 Ly$\alpha$ & $<0.041$ &  $<0.088$ \\
      \hline
      +eBOSS Ly$\alpha$ (shifted)&  $<0.039$ &  $<0.063$ \\%$0.019^{+0.043}_{-0.019}$ \\  
      \hline
     
    \end{tabular}
    }
    \caption{Comparison of constraints on $f_{\rm EDE}$ for the MCMC and profile likelihood analyses of the main datasets considered in this work. The MCMC constraints correspond to the one-tailed 95\% CL upper bound. The profile likelihood constraints are derived using the Feldman-Cousins prescription~\cite{Feldman:1997qc}, resulting in a 95\% CL upper bound for the ``shifted" eBOSS analysis and the XQ-100 analysis. Note that for the baseline and eBOSS analyses, the 95\% CL lower bound derived from the Feldman--Cousins prescription is non-zero. Consequently, we report the 95\% confidence interval centered at the minimum of the parabola for these datasets.}
   \label{tab:prof_like_table}
   \end{table}

To construct confidence intervals, we first fit the $\Delta \chi^2(f_{\rm EDE})$ values with a two-parameter parabolic model $\Delta\chi^2(f_{\rm EDE})=(f_{\rm EDE}-f_{\rm EDE}^0)^2/(\sigma^0_{f_{\rm EDE}})^2$ where $f_{\rm EDE}^0$ and $\sigma^0_{f_{\rm EDE}}$ characterize the center and width of the parabola, respectively. As seen in Fig.~\ref{fig:profile_likelihood}, the parabola provides an excellent fit to the $\Delta\chi^2(f_{\rm EDE})$ values for the dataset combinations analyzed in the main text. Since the profile likelihoods are close to the physical boundary $f_{\rm EDE}=0$, the Neyman prescription~\cite{Neyman} to determine the 68.27\% confidence interval (\emph{i.e.}, finding where the parabola intersects $\Delta\chi^2(f_{\rm EDE})=1$) is insufficient. Therefore, we use the Feldman--Cousins prescription~\cite{Feldman:1997qc} to determine confidence intervals.

The Feldman--Cousins prescription for a given parabolic fit relies on the likelihood ratio, denoted as 
\begin{equation}
    R(x)=\frac{\mathcal{L}(x|\mu)}{\mathcal{L}(x|\mu_{\rm best})},
\end{equation}
where $x$ ranges over all allowed values of $f_{\rm EDE}$. Here, $\mu=f_{\rm EDE}^0$ corresponds to the ``true" value of $f_{\rm EDE}$ obtained from the parabolic fit, while $\mu_{\rm best}$ represents the best-fit allowed value of $f_{\rm EDE}$ for a given $x$. Since $f_{\rm EDE}$ is non-negative, $\mu_{\rm best}=x$ if $x\geq 0$, otherwise $\mu_{\rm best}=0.$ To construct a confidence interval at confidence level $\alpha$ (\textit{e.g.}, $\alpha=0.95)$, we need to find an interval $[x_1, x_2]$ that satisfies $R(x_1)=R(x_2)$ and 
\begin{equation}
    \int_{x_1}^{x_2}\mathcal{L}(x|\mu)dx=\alpha.
\end{equation}
These intervals have been computed numerically and are tabulated in~\cite{Feldman:1997qc}.

In Table~\ref{tab:prof_like_table}, we compare the 95\% confidence intervals on $f_{\rm EDE}$ from the MCMC analyses with those from the profile likelihood using the Feldman--Cousins prescription. For the MCMC analyses, we report the one-tailed 95\% CL upper bound. For the profile likelihood analyses, we report a one-tailed 95\% CL upper bound if the lower bound is consistent with zero; otherwise, we report the two-tailed central confidence interval. Using the profile likelihood, we find $f_{\rm EDE}=0.069^{+0.074}_{-0.062}$ at 95\% CL for the baseline dataset, consistent with the value of $f_{\rm EDE}=0.072^{+0.071}_{-0.060}$ at 95\% CL from the \emph{Planck} + BOSS full-shape analysis presented in~\cite{Herold:2021ksg}. This significantly deviates from the MCMC-inferred bound of $f_{\rm EDE}<0.073$ at 95\% CL, indicating that the MCMC analysis of the baseline dataset is impacted by prior-volume effects. For the analyses including the Ly$\alpha$ forest measurements, the discrepancy between the MCMC and profile likelihood constraints is significantly less pronounced. Crucially, the 95\% CL upper bounds for the eBOSS and shifted eBOSS profile likelihood analyses (corresponding to the MCMC analyses shown in Fig.~2 of the main text) are $f_{\rm EDE}<0.039$ and $f_{\rm EDE}<0.063$, respectively, strongly excluding the $f_{\rm EDE}\approx0.1$ value necessary to resolve the Hubble tension. Unsurprisingly, the XQ-100 Ly$\alpha$ analysis is more prone to prior-volume effects than the eBOSS analysis, due to its weaker constraining power. However, incorporating the XQ-100 dataset still leads to a significant reduction in the profile likelihood upper bound on $f_{\rm EDE}$, from 0.143 to 0.088 at 95\% CL.

In conclusion, the profile likelihood analyses presented in this section demonstrate that the conclusions drawn from the MCMC analyses in the main text are not an artifact of prior-volume effects. While it is true that the profile likelihood analyses prefer a slightly higher value for the upper bound on $f_{\rm EDE}$ compared to the MCMC-derived bound, indicating that prior-volume effects do impact our MCMC analyses, the key takeaway is the remarkable reduction in the upper bound on $f_{\rm EDE}$ once the Ly$\alpha$ datasets are incorporated. This analysis supports the main text's conclusion that current Ly$\alpha$ forest observations impose significant constraints on canonical EDE cosmologies.  

\begin{table*}[!htbp]
    \centering
    {\scriptsize
    \begin{tabular}{ |c|c|c|c|c|c|c|c|c|c|c| }
     \hline
      & $f_{\rm EDE} $ &  $\log_{10}(z_c)$  &  $\theta_i$ & $\log(10^{10}A_s)$  &  $n_s$ &   $H_0$  & $\Omega_ch^2$  & $\Omega_bh^2$ & $\Omega_m$ & $\tau$ \\ 
      \hline 
      Baseline     &  $<0.073$ &  $3.67^{+0.24}_{-0.30}$ & $1.99^{+1.11}_{-0.36}$ & $3.047^{+0.014}_{-0.015}$ & $0.9705^{+0.0046}_{-0.0066}$ & $68.92^{+0.55}_{-0.91}$ & $0.1221^{+0.0013}_{-0.0031}$ & $0.0225^{+0.0002}_{-0.0002}$ & $0.3045^{+0.0053}_{-0.0058}$ & $0.0544^{+0.0068}_{-0.0068}$ \\ 
      \hline
       {} & 0.072 & $3.56$ &$2.73$ &$3.055$ &$0.9791$ &$70.33$ &$0.1267$ &$0.0226$ &$0.3019$ &$0.0537$ \\
       \hline
       {+SH0ES} & $0.096^{+0.032}_{-0.026}$ &$3.64^{+0.21}_{-0.16}$ & $2.55^{+0.39}_{-0.06}$ & $3.062^{+0.015}_{-0.015}$ & $0.9851^{+0.0065}_{-0.0063}$ & $71.40^{+0.91}_{-0.91}$ & $0.1287^{+0.0035}_{-0.0035}$ & $0.0228^{+0.0002}_{-0.0002}$ & $0.2972^{+0.0049}_{-0.0049}$ & $0.0566^{+0.0073}_{-0.0072}$ \\
       \hline
       {} & $0.114$ &$3.57$ &$2.76$ &$3.065$ &$0.9878$ &$72.01$ &$0.1311$ &$0.0227$ &$0.2966$ &$0.0554$\\
      \hline
      {+eBOSS Ly$\alpha$}     & $<0.028$   &$3.52^{+0.78}_{-0.52}$ & $1.68^{+1.42}_{-1.58}$ & $3.038^{+0.013}_{-0.013}$ & $0.9549^{+0.0039}_{-0.0035}$ & $67.88^{+0.43}_{-0.46}$ & $0.1211^{+0.0011}_{-0.0011}$ & $0.0223^{+0.0002}_{-0.0002}$ & $0.3114^{+0.0056}_{-0.0057}$ & $0.0503^{+0.0066}_{-0.0066}$  \\
      \hline
       {} & 0.021 &$3.03$ &$3.10$ &$3.036$ &$0.9510$ &$67.81$ &$0.1213$ &$0.0221$ &$0.3118$ &$0.0510$ \\
       \hline
      {+eBOSS+SH0ES}  &  $<0.035$ &$3.48^{+0.82}_{-0.48}$ & $1.75^{+1.35}_{-0.47}$ & $3.044^{+0.013}_{-0.014}$ & $0.9574^{+0.0044}_{-0.0037}$ & $68.69^{+0.42}_{-0.41}$ & $0.1202^{+0.0010}_{-0.0014}$ & $0.0224^{+0.0002}_{-0.0002}$ & $0.3022^{+0.0046}_{-0.0052}$ & $0.0546^{+0.0065}_{-0.0071}$\\ 
      \hline
      {} & 0.026 & $3.06$ &$3.10$ &$3.042$ &$0.9532$ &$68.66$ &$0.1203$ &$0.0222$ &$0.3022$ &$0.0546$ \\
      \hline
      {+XQ-100 Ly$\alpha$}  &  $<0.041$ & $3.61^{+0.28}_{-0.42}$ & $1.76^{+1.34}_{-0.48}$ & $3.050^{+0.014}_{-0.014}$ & $0.9646^{+0.0041}_{-0.0050}$ & $68.28^{+0.47}_{-0.66}$ & $0.1216^{+0.0011}_{-0.0018}$ & $0.0224^{+0.0002}_{-0.0002}$ & $0.3089^{+0.0057}_{-0.0059}$ & $0.0556^{+0.0066}_{-0.0072}$  \\ 
      \hline
      {} & 0.022 &$3.53$ &$2.75$ &$3.051$ &$0.9666$ &$68.46$ &$0.1223$ &$0.0224$ &$0.3087$ &$0.0552$ \\
      \hline
      {+XQ-100+SH0ES}  & $0.060^{+0.025}_{-0.028}$  &$3.55^{+0.06}_{-0.15}$ & $2.07^{+0.65}_{-1.97}$ & $3.063^{+0.015}_{-0.015}$ & $0.9750^{+0.0054}_{-0.0060}$ & $70.25^{+0.84}_{-0.84}$ & $0.1256^{+0.0032}_{-0.0034}$ & $0.0227^{+0.0002}_{-0.0002}$ & $0.3005^{+0.0052}_{-0.0052}$ & $0.0581^{+0.0070}_{-0.0078}$ \\ 
      \hline
        {} & $0.092$ & $3.54$ &$2.78$ &$3.065$ &$0.9788$ &$71.01$ &$0.1289$ &$0.0226$ &$0.3005$ &$0.0564$ \\
        \hline
        {+eBOSS (shfited)}  & $<0.039$ & $3.69^{+0.61}_{-0.19}$ & $1.81^{+1.29}_{-0.50}$ & $3.044^{+0.014}_{-0.014}$ & $0.9668^{+0.0035}_{-0.0036}$ & $68.49^{+0.46}_{-0.55}$ & $0.1209^{+0.0011}_{-0.0017}$ & $0.0225^{+0.0002}_{-0.0002}$ & $0.3057^{+0.0053}_{-0.0054}$ & $0.0535^{+0.0069}_{-0.0068}$ \\
      \hline
        {} & $0.019$ &$3.55$ &$2.72$ &$3.045$ &$0.9678$ &$68.57$ &$0.1215$ &$0.0224$ &$0.3062$ &$0.0531$ \\
     \hline
    \end{tabular}
    }
   \caption{Marginalized constraints on cosmological parameters for EDE ($n=3$) from the datasets considered in this work. For each dataset, we report the posterior mean and the 68\% CL upper and lower limits for all parameters that are detected at $>2\sigma$, otherwise we report the 95\% CL upper limits.  Maximum \textit{a posteriori} values for each dataset are shown in the following row. $H_0$ values are in [km/s/Mpc].} 
   \label{tab:posterior_full_supp_material}
\end{table*}

\subsection{$\Lambda$CDM results}

In this section, we analyze the baseline + eBOSS Ly$\alpha$ and baseline + XQ-100 Ly$\alpha$ datasets assuming a $\Lambda$CDM cosmology with massless neutrinos. Fig.~\ref{fig:ly_alpha_LCDM_constraints} shows the marginalized posteriors for the six $\Lambda$CDM parameters for the baseline dataset, as well for combinations of the baseline dataset with the eBOSS and XQ-100 Ly$\alpha$ datasets. Similar to our findings in the EDE analysis, the main impact of including Ly$\alpha$ data is to drive the posterior towards lower values of $n_s.$

The top right panel presents constraints on the effective Ly$\alpha$ parameters $n_L$ and $\Delta_L^2$ for our main analyses, alongside the Ly$\alpha$ likelihoods for the eBOSS (blue dashed) and XQ-100 (red dashed) Ly$\alpha$ datasets. As in the EDE analysis, we find significant disagreement between the grey contour and the dashed likelihoods. The tension between \emph{Planck} and eBOSS Ly$\alpha$ is discussed in some detail in \cite{Palanque-Delabrouille:2019iyz}, where it is identified as a mild tension in both $n_s$ and $\Omega_m.$ The tension is particularly apparent in the $\Delta_L^2$ -- $n_L$ plane since these parameters are sensitive to both $n_s$ and $\Omega_m$.  Quantifying this disagreement using various tension metrics~(\textit{e.g.},~\cite{Raveri:2018wln,Nicola:2018rcd,Handley:2019wlz}) would be of interest.  Resolving this tension could indeed necessitate new physics~(\textit{e.g.},~\cite{Hooper:2021rjc,Hooper:2022byl}), but in the exact opposite direction to that introduced by EDE.

\subsection{Origin of  $\log_{10}(z_c)$ bimodality in eBOSS analysis}

In this section, we discuss the physical origin of the bimodality in the $\log_{10}(z_c)$ posterior for the baseline + eBOSS Ly$\alpha$ analysis shown in Fig.~\textcolor{Maroon}{2} of the main text. As shown in Fig.~2 of \cite{Poulin:2018cxd}, for $3 \lesssim \log_{10}(z_c) \lesssim 3.3$, the ratio of the comoving sound horizon at last scattering, $r_s$, to the Silk damping scale, $r_D$, increases compared to its $\Lambda$CDM value.\footnote{For $\log_{10}(z_c) \gtrsim 3.3$, $r_s/r_D$ decreases in EDE compared to $\Lambda$CDM.}  Thus, $\theta_s/\theta_D \equiv (r_s/d_A^\star)/(r_D/d_A^\star)$ also increases, where $d_A^\star$ is the comoving angular diameter distance to recombination. Since the angular size of the sound horizon $\theta_s$ is effectively fixed by observations, the angular Silk damping scale $\theta_D$ thus decreases (\textit{i.e.}, $\ell_D$ increases). In the high-$\ell$ CMB, there is thus less damping at a given multipole, and this is compensated by a decrease in $n_s$ to maintain the fit to the CMB data.  This explains the anti-correlation between $n_s$ and $f_{\rm EDE}$ seen in our baseline + eBOSS Ly$\alpha$ analysis; only the low-$z_c$ region can (partially) accomodate the low values of $n_s$ preferred by the eBOSS Ly$\alpha$ data. This feature could be a useful hint toward constructing models that can resolve the Hubble tension and accommodate Ly$\alpha$ forest data.

\subsection{Cosmological parameters used in Figure 1}

In this section, we discuss the cosmological parameters used to estimate the best-fit linear power spectra for the eBOSS and XQ-100 datasets shown in Fig.~\textcolor{Maroon}{1} of the main text. For the eBOSS cosmology, we use the best-fit cosmological parameters from the second row of Table 7 of~\cite{Chabanier:2018rga}. These parameters are $\Omega_m=0.269,$ $\sigma_8=0.820$, $n_s=0.955$, $h=0.671$. Additionally, we fix $\Omega_b=0.05$ to be consistent with the central values of the compressed eBOSS Ly$\alpha$ likelihood, which are $\Delta^2_L=-2.34$ and $n_L=0.31$. For the XQ-100 analysis, we use the best-fit parameters from the analysis in Appendix A of~\cite{Esposito:2022plo}. This analysis varies only $\sigma_8$, $n_{\rm eff}$, and astrophysical IGM parameters. Based on the simulations used in~\cite{Esposito:2022plo}, the best-fit parameters correspond to $A_s=2.663\times 10^{-09}$, $n_s=0.875$, $\Omega_m=0.308$, $\Omega_b=0.0482$, and $h=0.678$. The XQ-100 cosmology also assumes a single species of massive neutrinos with $\sum m_\nu=0.06$ eV.

\subsection{Full table of parameter constraints}
Table~\ref{tab:posterior_full_supp_material} contains the full set of marginalized constraints on cosmological parameters for EDE ($n=3$) using the datasets considered in our main analysis. 
%TC:endignore

\end{document}